 \documentclass[final,authoryear,3p,11pt]{elsarticle}

\usepackage{graphicx}
\usepackage{epstopdf}
\usepackage{booktabs}
\usepackage{natbib}
\usepackage{fancyhdr}
\usepackage{amsmath}
\usepackage{amsfonts}
\usepackage{multirow}
\usepackage{lscape}
\usepackage{color}
\usepackage{hyperref}
\usepackage{color,soul}
\sethlcolor{green}
\usepackage{ifthen}
\usepackage[latin1]{inputenc}
\usepackage{titletoc}
\usepackage{eso-pic}
\definecolor{123}{rgb}{.9,.9,.9}
\hypersetup{
colorlinks=false,
citecolor=blue,
linkbordercolor={1 1 1}, 
citebordercolor={1 1 1},
urlbordercolor={1 1 1}
}

\usepackage{amssymb}
 \usepackage{amsthm}
 \usepackage{xcolor}

\newtheorem{prop}{Proposition}

\newtheorem{def1}{Definition}

\newcommand{\ra}[1]{\renewcommand{\arraystretch}{#1}}

\newcommand{\EL}{\ensuremath{L^{2}(\mathbb  R)}}

\def \ZN {\mathbb Z_N}
\def \R {\mathbb  R}

\begin{document}

\begin{frontmatter}

%
%
\title{Realized wavelet-based estimation of integrated variance and jumps in the presence of noise \tnoteref{label1} }

%
%
\author[ies,utia]{Jozef Barunik\corref{cor2} } \ead{barunik@utia.cas.cz}
\author[ies,utia]{Lukas Vacha}
\address[ies]{Institute of Economic Studies, Charles University, Opletalova 21, 110 00, Prague,  CR}
\address[utia]{Institute of Information Theory and Automation, Academy of Sciences of the Czech Republic, Pod Vodarenskou Vezi 4, 182 00, Prague, Czech Republic}

\tnotetext[label1]{We are grateful to David Veredas and Karel Najzar for many useful comments and suggestions. We are also grateful to seminar participants at the Modeling High Frequency Data in Finance 3 in New York (July 2011) and Computational and Financial Econometrics in London (December 2011) for many useful discussions. The support from the Czech Science Foundation under the 402/09/H045 project is gratefully acknowledged.}

\begin{abstract}
We introduce wavelet-based methodology for estimation of realized variance allowing its measurement in the time-frequency domain. Using smooth wavelets and Maximum Overlap Discrete Wavelet Transform, we allow for the decomposition of the realized variance into several investment horizons and jumps. Basing our estimator in the two-scale realized variance framework, we are able to utilize all available data and get feasible estimator in the presence of microstructure noise as well. The estimator is tested in a large numerical study of the finite sample performance and is compared to other popular realized variation estimators. We use different simulation settings with changing noise as well as jump level in different price processes including long memory fractional stochastic volatility model. The results reveal that our wavelet-based estimator is able to estimate and forecast the realized measures with the greatest precision. Our time-frequency estimators not only produce feasible estimates, but also decompose the realized variation into arbitrarily chosen investment horizons. We apply it to study the volatility of forex futures during the recent crisis at several investment horizons and obtain the results which provide us with better understanding of the volatility dynamics. 

\end{abstract}
\end{frontmatter}


\section{Introduction}

Volatility of asset returns has become one of the primary concerns in financial econometrics research over the past decades. One of the main improvements in this area has been introduction of high frequency data into the volatility estimation. The most popular \textit{Realized Volatility} approach was pioneering work which took advantage of the data in a nonparametric fashion, but as both theoretical insights and data availability have grown rapidly in the past decade, this research line has brought great improvements in the volatility estimation and forecasting. While simple realized volatility estimator of \cite{abdl2003} is still the most popular among researchers due to its simplicity, literature shown that it is important to improve the estimation as microstructure noise \citep{zhang2005,hansenlunde2006,bandirussel2006a,barndorff2008} and relatively frequent jumps \citep{barndorff2006,huang2005,abd2007,ads2009,sahaliajacod2009,Mancini2009} play an important role in the price process. 

Our motivation is to bring the time-frequency insight into volatility estimation for the first time while most time series models are set in the time domain or frequency domain separately. This is enabled by the use of the wavelet transform. It is a logical step to take, as the stock markets are believed to be driven by heterogeneous investment horizons. In our work, we ask if wavelet decomposition can improve our understanding of volatility series and hence improve volatility forecasting and risk management.

On the theoretical side, wavelets can be easily embedded into stochastic processes, as shown by \cite{antoniou1999} and we can conveniently use them in the quadratic variation estimation. Several attempts to use wavelets in the estimation of realized variation have emerged in the past few years. \cite{hoglunde2003} were the first to suggest a wavelet estimator of realized variance. \cite{capobianco2004}, for example, proposes to use a wavelet transform as a comparable estimator of quadratic variation. \cite{subbotin2008} uses wavelets to decompose volatility into a multi-horizon scale. Next, \cite{nielsenfrederiksen2008} compare the finite sample properties of three integrated variance estimators, i.e., realized variance, Fourier and wavelet estimators. They consider several processes generating time series with a long memory, jump processes as well as bid-ask bounce. \cite{gencay2010} mention the possible use of wavelet multiresolution analysis to decompose realized variance in their paper, while they concentrate on developing much more complicated structures of variance modeling in different regimes through wavelet-domain hidden Markov models. Work of \cite{fanwang2008} fully completes the current literature on using wavelets in realized variation measurement. Authors utilize wavelets to build the methodology for the estimation of jumps. Finally,  \cite{mancino2008}, \cite{olhede2009} propose estimators based on the Fourier transform. While the idea is very similar, this approach leads to realized volatility measurement in the frequency domain solely.

In our work, we revisit and extend these results in several ways. Instead of using the Discrete Wavelet Transform we use the Maximum Overlap Discrete Wavelet Transform, which is a more efficient estimator and is not restricted to sample sizes that are powers of two. We also use smooth wavelets, specifically the Daubechies family of wavelets instead of the Haar type.

An important contribution of this paper is that we allow for decomposition of the realized variance into several investment horizons. Basing our estimator in the two-scale realized variance framework, we are able to utilize all available data and get feasible estimator in the presence of microstructure noise as well as jumps. To study the finite sample behavior of the estimator, we run a large numerical study using several price generating processes including a long memory fractional stochastic volatility model. We use several different jumps and noise levels to compare our estimator to other commonly used estimators, namely realized variance, bipower variation, realized kernels and two-scale realized volatility. Next, we also run a simulation study comparing the forecasting ability of the estimators. The results suggest that under various settings our wavelet-based estimator proves to have the lowest forecast bias.

In the final part of the paper, we apply the wavelet-based estimator to the modeling of currency futures volatility. By studying the statistical properties of unconditional daily log-return distributions standardized by volatility estimated using the different estimators we find that standardization by our wavelet-based estimator brings the returns close to the Gaussian normal distribution. The differences to other estimators are quite large, as we find that the average volatility estimated using our wavelet-based theory is 6.34\% lower than the volatility estimated with the standard estimator. More importantly, we study the volatility decomposed to several investment horizons and jumps on the recent data covering financial crisis.

Organization of the paper is as follows. The second Section introduces estimators of integrated variance commonly used in the literature which will be used as a benchmark in our study. The third section introduces wavelet decomposition of integrated variance and  derives wavelet-based realized variance estimator and its properties. The fourth Section tests the theory in a numerical study and compares the small sample behavior of the wavelet-based estimator with other popular estimators, while assuming different processes driving the stock market with different amounts of noise and jumps. Specifically, we consider jump-diffusion stochastic volatility and fractional stochastic volatility. The Section concludes with a numerical study assessing the forecasting performance of the estimators. The last Section applies the presented theory and decomposes the empirical volatility of forex stock markets.




\section{Estimation of integrated variance}

Consider a univariate risky logarithmic asset price process $p_t$ defined on a complete probability space $ (\Omega,\mathcal{F},\mathbb{P})$. The price process evolves in continuous time over the interval $\left[0,T\right]$, where $T$ is a finite positive integer. Further, consider the natural information filtration, an increasing family of $\sigma$-fields $(\mathcal{F}_t)_{t\in \left[0,T\right]}\subseteq \mathcal{F}$, which satisfies the usual conditions. Following \cite{abdl2003}, we define the continuously compounded asset return over the $\left[t-h,t\right]$ time interval, $0 \le h\le t \le T$, by $r_{t,h}=p_t-p_{t-h}$. Instantaneous return can be uniquely decomposed into a predictable and integrable mean (expected return) component and a local martingale innovation (e.g. \citealp{protter}). For any univariate, square-integrable, continuous sample path, logarithmic price process $\left(p_t\right)_{t\in\left[0,T\right]}$ which is not locally riskless, there exists a representation such that over $\left[t-h,t\right]$, for $0\le h \le t \le T$
\begin{equation}
\label{representation}
r_{t,h}=\int_{t-h}^t \mu_s ds+\int_{t-h}^t \sigma_s dW_s,
\end{equation}
where $\mu_s$ is an integrable, predictable and finite-variation stochastic process, $\sigma_s$ is a strictly positive c\`adl\`ag stochastic process satisfying
\[
P\left[\int_{t-h}^t \sigma_s^2 ds<\infty \right]=1,
\]
and $W_t$ is a standard Brownian motion.

In the observed data the logarithmic asset price is latent as it is contamined with microstructure noise and moreover contain jumps. Thus we assume that the latent price process follows a standard jump-diffusion process and is contamined with microstructure noise. 

Let  $\left(y_t\right)_{t\in\left[0,T\right]}$  be the observed log prices, which will be equal to the latent, so-called ``true log-price process"
\begin{equation}
dp_t=\mu_t dt+\sigma_t dW_t+\xi_t dq_t,
\end{equation}
and will contain microstructure noise $\epsilon_t$
 \begin{equation}
 \label{eqnoise}
 y_t=p_t+\epsilon_t,
 \end{equation}
 where $\epsilon_t$ is zero mean $i.i.d.$ noise with variance $\eta^2$, $q$ is a Poisson process uncorrelated with $W$ and governed by the constant jump intensity $\lambda$. The magnitude of the jump in the return process is controlled by factor $\xi_t \sim N(\bar\xi,\sigma^2_{\xi})$.

Quadratic return variation over the $\left[t-h,t\right]$ time interval, $0 \le h \le t \le T$, associated with $p_t$,
\begin{equation}
\label{waveqvjump}
QV_{t,h}=\underbrace{\int_{t-h}^t \sigma_s^2 ds}_{\mbox{$IV_{t,h}$}}+ \underbrace{\sum_{t-h \le l \le t} J_l^2}_{\mbox{$JV_{t,h}$}} 
\end{equation}
can be naturally decomposed into two parts: integrated variance of the latent price process,  $IV_{t,h}$ and jump variation $JV_{t,h}$. As detailed by \cite{abdl2003}, quadratic variation is a natural measure of variability in the logarithmic price.

A simple consistent estimator of the overall quadratic variation under the assumption of zero noise contamination in the price process is provided by the well-known realized variance, introduced by \cite{ab98}. The realized variance over $\left[t-h,t\right]$, for $0\le h \le t\le T$, is defined by
\begin{equation}
\label{rv}
\widehat{RV}_{t,h}=\sum_{i=1}^N r_{t-h+\left(\frac{i}{N}\right)h}^2,
\end{equation}
where $N$ is the number of observations in $\left[t-h,t\right]$ and $r_{t-h+\left(\frac{i}{N}\right)h}$ is $i-$th intraday return in the $\left[t-h,t\right]$ interval. $\widehat{RV}_{t,h}\overset{p}{\to}IV_{t,h}+JV_{t,h}$ as $N\rightarrow\infty$ \citep{ab98, abdl2001,abdl2003, barndorff2001,barndorff2002a,barndorff2002}. In the subsequent literature, \cite{zhang2005,hansenlunde2006,bandirussel2006a,barndorff2008,barndorff2006,huang2005,abd2007,ads2009,sahaliajacod2009,Mancini2009} show that it is important to account for the microstructure noise and jumps. In our study, we use these estimators for comparison to our wavelet-based approach, thus we introduce them in the following Section. 

\subsection{Effect of microstructure noise}

\cite{zhang2005} propose the solution to the noise contamination by introducing so-called two-scale realized volatility (TSRV henceforth) estimator. Authors propose a methodology for measurement of realized variance utilizing all of the available data using an idea of precise bias estimation. The two-scale realized variation over $\left[t-h,t\right]$, for $0\le h \le t\le T$, is measured by
\begin{equation}
\label{tsrv}
\widehat{RV}_{t,h}^{(TSRV)}=\underbrace{\widehat{RV}_{t,h}^{(average)}}_{\mbox{\footnotesize slow time scale}}-\frac{\bar{N}}{N} \underbrace{\widehat{RV}_{t,h}^{(all)}}_{\mbox{\footnotesize fast time scale}},
\end{equation}
where $\widehat{RV}_{t,h}^{(all)}$ is computed using Eq. (\ref{rv}) on all available data and $\widehat{RV}_{t,h}^{(average)}$ is constructed by averaging the estimators $\widehat{RV}_{t,h}^{(k)}$ obtained on $K$ grids of average size $\bar{N}=N/K$ as:
\begin{equation}
\widehat{RV}_{t,h}^{(average)}=\frac{1}{K}\sum_{k=1}^K \widehat{RV}_{t,h}^{(k)}.
\end{equation}
In computing the TSRV, we have to first partition the original grid of observation times, $G=\{t_0,\dots,t_N\}$, into subsamples $G^{(k)}$, $k=1,\dots,K$, where $N/K\rightarrow \infty$ as $N\rightarrow \infty$. For example, $G^{(1)}$ will start at the first observation and take an observation every 5 minutes, $G^{(2)}$ will start at the second observation and take an observation every 5 minutes, etc. Finally, we  average these estimators through the subsamples, so we average the variation of the estimator as well. $\widehat{RV}_{t,h}^{(TSRV)}$ provides the first consistent estimator of the quadratic variation of $p_t$ with rate of convergence $N^{-1/6}$. \cite{zhang2005} also provide the distribution theory as well as theory for optimal choice of $K$ grids, $K^*=c N^{2/3}$, where the constant $c$ can be set to minimize the total asymptotic variance.

Another estimator, which is able to deal with the noise and which we use for the comparison in our study is the realized kernels (RK) estimator introduced by \cite{barndorff2008}. The realized kernel variance estimator is defined by
 \begin{equation}
 \label{rk}
  \widehat{RV}_{t,h}^{(RK)}=\gamma_{t,h,0} +\sum_{\eta=1}^H k \left(\frac{\eta-1}{H}\right) (\gamma_{t,h,\eta}+\gamma_{t,h,-\eta}),
   \end{equation}
with $\gamma_{t,h,\eta}=\sum_{i=1}^N r_{t-h+\left(\frac{i}{N}\right)h} r_{t-h+\left(\frac{i-\eta}{N}\right)h}$ denoting the $\eta$-th realized autocovariance with $\eta=-H,\dots,-1,0,1,\dots,H$ and $k(.)$ denotes the kernel function. Please note that for $\eta=0$, $\gamma_{t,h,\eta}=\gamma_{t,h,0}=\widehat{RV}_{t,h}$ is estimate of the realized variance from Eq. (\ref{rv}). For the estimator to work, we need to choose the kernel function $k(.)$. In our study, we will focus on the Parzen kernel because it satisfies the smoothness conditions, $k'(0)=k'(1)=0$ and is guaranteed to produce a non-negative estimate. The Parzen kernel function is given by
\begin{equation}
k(x)=
\left\{
\begin{array}{ll}
1-6x^2+6x^3 & 0 \le x \le 1/2 \\
2(1-x)^3 & 1/2 \le x \le 1. \\
0 & x>1
\end{array}
\right.
\end{equation}
We should note that the realized kernel estimator is computed without accounting for end effects, i.e. replacing the first and the last observation by local averages to eliminate the corresponding noise components (so-called ``jittering"). \cite{barndorff2008} argue that these effects are important theoretically, but are negligible practically.

\subsection{Effect of jumps}

By introducing the TSRV and the RK estimators, we will have benchmark estimators which are able to consistently estimate the quadratic variation from noisy observations. Still, we are interested to decompose quadratic variation into the integrated variance and jump variation component. \cite{barndorff2004,barndorff2006} develop a powerful and complete way of detecting the presence of jumps in high-frequency data. The basic idea is to compare two measures of the integrated variance, one containing the jump variation and the other being robust to jumps and hence containing only the integrated variation part. In our work, we use the \cite{ABH2011} adjustment of the original \cite{barndorff2004} estimator, which helps render it robust to certain types of microstructure noise. The bipower variation over $[t-h,t]$, for $0 \le h \le t \le T$, is defined by
 \begin{equation}
 \label{bv}
   \widehat{RV}^{(BV)}_{t,h}=\mu_1^{-2} \frac{N}{N-2} \sum_{i=3}^{N} |r_{t-h+\left(\frac{i-2}{N}\right)h}|.|r_{t-h+\left(\frac{i}{N}\right)h}|,
   \end{equation}
   where $\mu_a=\pi/2=E(|Z|^a),$ and $Z\sim N(0,1)$, $a \ge0$ and $ \widehat{RV}^{(BV)}_{t,h} \overset{p}{\to} \int_{t-h}^t \sigma_s^2 ds$.
Thus $ \widehat{RV}^{(BV)}_{t,h}$ provides a consistent estimator of the integrated variance. While $ \widehat{RV}^{(sparse)}_{t,h}$ provides a consistent estimator of the quadratic variation, the jump variation can be estimated consistently as the difference between the realized variance and the realized bipower variation: 
\begin{equation}
\left(\widehat{RV}^{(sparse)}_{t,h}-\widehat{RV}^{(BV)}_{t,h}\right)\overset{p}{\to}JV_{t,h}.
\end{equation}
Under the assumption of no jump and some other regularity conditions, \cite{barndorff2006} provided the joint asymptotic distribution of the jump variation. Under the null hypothesis of no within-day jumps,
 \begin{equation}
   Z_{t,h}=\frac{\frac{\widehat{RV}^{(sparse)}_{t,h}-\widehat{RV}^{(BV)}_{t,h}}{\widehat{RV}^{(sparse)}_{t,h}}}{\sqrt{((\frac{\pi}{2})^2+\pi-5)\frac{1}{n} \max \left( 1,\frac{\widehat{TQ}_{t,h}}{\left(\widehat{RV}^{(BV)}_{t,h} \right)^2} \right)}},
   \end{equation}
   where $\widehat{TQ}_{t,h}=N \mu_{4/3}^{-3}(\frac{N}{n-4}) \sum_{j=5}^{N} |r_{t-h+\left(\frac{i-4}{N}\right)h}|^{4/3}|r_{t-h+\left(\frac{i-2}{N}\right)h}|^{4/3} |r_{t-h+\left(\frac{i-2}{N}\right)h}|^{4/3}$ is asymptotically standard normally distributed. Using this theory, the contribution of the jump variation to the quadratic variation of the price process  is measured by

\begin{equation}
\label{bvjump1}
\widehat{J}_{t,h}=I_{Z_{t,h}>\Phi_{\alpha}} \left (\widehat{RV}^{(sparse)}_{t,h}-\widehat{RV}^{(BV)}_{t,h} \right),
\end{equation}
where $I_{Z_{t,h}>\Phi_{\alpha}}$ denotes the indicator function and $\Phi_{\alpha}$ refers to the chosen critical value from the standard normal distribution. The measure of integrated variance is defined as
\begin{equation}
\label{bvc}
\widehat{C}_{t,h}=I_{Z_{t,h} \le \Phi_{\alpha}} \widehat{RV}^{(sparse)}_{t,h}+I_{Z_{t,h}> \Phi_{\alpha}}\widehat{RV}^{(BV)}_{t,h},
\end{equation}
ensuring that the jump measure and the continuous part add up to the estimated variance without jumps.

We use the described jump detection methodology as the benchmark and we focus on wavelet methods for detecting jumps in the data, as described in the following sections.


\section{Wavelet decomposition of integrated variance}

While most realized variance estimators are naturally set in the time domain, or frequency domain separately, wavelet transform help us to enrich the analysis of realized variance in the time-frequency domain.  It is a logical step to take, as the stock markets are believed to be driven by heterogeneous investment horizons, so volatility dynamics should be understood not only in time but at investment horizons as well. We will introduce general ideas of constructing the estimators here.

\subsection{Decomposition of quadratic return variation with wavelet transform}

The quadratic variation can be decomposed using the continuous wavelet transform (CWT): 
\begin{equation}
\label{qvw}
QV_{t,h}=\underbrace{\frac{2}{C_\psi}\int_{t-h}^t \int_{0}^{\infty} \int_{\mathbb{R}}\psi_{j,k}(s)\langle\psi_{j,k},\sigma_s^2 \rangle dk \frac{1}{j^2}djds}_{\mbox{$IV_{t,h}$}} + \underbrace{\sum_{t-h \le l \le t} J_l^2}_{\mbox{$JV_{t,h}$}},
\end{equation}
where
\begin{equation}
\langle\psi_{j,k} ,\sigma_s^2 \rangle=\vert j \vert ^{-1/2}\int_{\mathbb{R}}\overline{\psi\left(\frac{s-k}{j}\right)} \sigma_s^2(s) ds
\end{equation}
Eq.(\ref{qvw}) decomposes the quadratic variation both in time and frequency. By decomposition in the frequency domain we obtain $j$ components representing scales which can be viewed as investment or  trading horizons. For more details about wavelet decomposition, consult \ref{Appwaveintro}. Further, Eq.(\ref{qvw}), allows to define a model-free measure of the integrated variation in analogy to the simple realized variance estimator.

 The continuous wavelet transform is a very important concept which helps us with the derivation of theoretical behavior on the time-scale space. Since we work with real data, we need some form of sampling to compute the estimators, i.e., we have to use a suitable form of discretization. We use the maximal overlap discrete wavelet transform (MODWT), which is a special form of discrete wavelet transformation, thus we restrict the scale $j$ and the translation $k$ parameters to integers only. Again, we keep the technical details about the MODWT in \ref{modwt}.

\subsection{Time-frequency decomposition of a stochastic process}

For our analysis, it is important to show that we are able to decompose the energy of a stochastic process on a scale-by-scale basis, i.e., we can obtain the energy contribution of every level $j$, with the maximum level of decomposition $J^m\le\log_2 N$. The (total) variance of the intraday returns $r_{t-h+\left(\frac{i}{N}\right)h}$ for $i=1,\ldots,N$ in the $\left[t-h,t\right]$ interval can be decomposed on a scale-by-scale basis $J^m\le\log_2 N$ so that 
\begin{equation}
\label{energydec}
\|\mathbf{r}\|^2=\sum_{j=1}^{J^m} \|\mathbf{W}_j\|^2 +\|\mathbf{V}_{J^m}\|^2
\end{equation}
where $\|\mathbf{r} \| ^2=\sum_{i=1}^{N}r_{t-h+\left(\frac{i}{N}\right)h}^2$,  $\|\mathbf{W}_j \| ^2=\sum_{i=1}^{N}W_{j,i}^2$, $\|\mathbf{V}_{J^m} \| ^2=\sum_{i=1}^{N}V_{J^m,i}^2\,$  and $\mathbf{W}_j$ and $\mathbf{V}_j$ are $N$ dimensional vectors of the $j$-th level MODWT wavelet and scaling coefficients. 

The Proof of the energy decomposition can be found in \cite{PercivalMofjeld1997}. It is central to derivation of wavelet-based estimators of integrated variance. It is worth noting that the squared norm $\|.\|$ is similar to the realized measure discussed in the preceding sections. For example, in the case of the realized variance estimator (RV) the variance decomposition can reveal the contributions of particular scales to the overall energy, hence we can see what form this realized measure takes. This will be introduced in the next paragraphs.

For simplicity in notation let us define a vector $\mathbf{\mathcal{W}}$ that consists of $J^m+1$ and $N-$dimensional subvectors, where the first $J^m$ subvectors are the MODWT wavelet coefficients at levels $j=1,...,J^m$ and the last subvector consists of the MODWT scaling coefficients at level $J^m$:
\begin{equation}
\mathcal{W}=\left( \mathbf{W}_1\mathbf{W}_2,\ldots,\mathbf{W}_{J^m}, \mathbf{V}_{J^m} \right)^T,
\end{equation}
i.e., for Equation (\ref{energydec}) the following holds:
\begin{equation}
\label{defW}
\|\mathbf{r}\|^2=\sum_{j=1}^{J^m} \|\mathbf{W}_j\|^2 +\|\mathbf{V}_{J^m}\|^2=\sum_{j=1}^{J^m+1}\|\mathcal{W}_j\|^2
\end{equation}

\subsection{Wavelet-based realized variance estimator}
Now we can return to the estimation of the realized variance and propose its wavelet-based estimator. The wavelet-based realized variance over $\left[t-h,t\right]$, for $0 \le h \le t \le T$, is defined by
\begin{equation}
\label{wrv}
\widehat{RV}_{t,h}^{(WRV)}=\sum_{j=1}^{J^m+1} \sum_{k=1}^N \mathcal{W}_{j,t-h+{\frac{k}{N}h}}^2,
\end{equation}
where $N$ is the number of intraday observations in $\left[t-h,t\right]$ and $J^m$ is the number of scales we consider. $\mathcal{W}_{j,t-h+{\frac{k}{N}h}}$ are the MODWT coefficients defined in Eq. (\ref{defW}) on returns data $r_{t,h}$ on components $j=1,\dots,J^m+1$, where $J^m \le\log_2 N$. This result comes readily from the results in the previous paragraphs. Using \cite{PercivalMofjeld1997} we can write that $\sum_{i=1}^{N} r_{t-h+{\frac{i}{N}h}}^2=\sum_{j=1}^{J^m+1} \sum_{k=1}^N \mathcal{W}_{j,t-h+{\frac{k}{N}h}}^2$ and thus we have readily that $\widehat{RV}_{t,h}=\widehat{RV}_{t,h}^{(WRV)}$. Moreover, $\widehat{RV}_{t,h}^{(WRV)}$ estimator takes asymptotic properties of $\widehat{RV}_{t,h}$ and converges in probability to quadratic variation
\begin{equation}
\widehat{RV}_{t,h}^{(WRV)} \overset{p}{\to}QV_{t,h}.
\end{equation}
The wavelet realized variance estimator in fact only decomposes the realized variance. Thus with increasing sampling frequency $N\to\infty$ it is an infeasible estimator of the quadratic variation in the presence of noise in the data. In the following section, we will introduce the concept of treating jumps using wavelets and finally propose the estimator, which will be able to estimate jumps consistently. Inheriting the structure of the TSRV our estimator will also utilize all the available data and will be feasible estimator of integrated variance under the microstructure noise.

\subsection{Realized jump estimation using wavelets \label{wavejumpsuni}}
Wavelets can be utilized for estimating jumps and separating integrated variance from jump variation. The sample path of $p_t$ has a finite number of jumps $(a.s.)$. Following the theoretical results of \cite{wang95} on the wavelet jump detection of the deterministic functions with $i.i.d.$ additive noise $\epsilon_t$, we use the MODWT as the discretized version of the continuous wavelet transform. Unlike the ordinary DWT, the MODWT is not restricted to a dyadic sample length. For the estimation of jump location we use the universal threshold \citep{donoho} on the first level wavelet coefficients of $y_t$ over $[t-h,t]$, $\mathcal{W}_{1,k}$. If for some $\mathcal{W}_{1,k}$
    \begin{equation}
    |\mathcal{W}_{1,k}| > d \sqrt{2 \log N},
   \end{equation}
   then $\hat{\tau}_l=\{k\}$ is the estimated jump location with size $\bar{y}_{\hat{\tau}_{l+}}-\bar{y}_{\hat{\tau}_{l-}}$ (averages over $[\hat{\tau}_l,\hat{\tau}_l+\delta_n]$ and $[\hat{\tau}_l,\hat{\tau}_l-\delta_n]$, respectively, with $\delta_n > 0$ being the small neighborhood of the estimated jump location\footnote{Due to the nature of the MODWT filters, we need to correct the position of the wavelet coefficient to get the precise position of the jump. For more details see \cite{PercivalMofjeld1997}.} $\hat{\tau}_l \pm \delta_n$) and where $d$ is median absolute deviation estimator defined as $(2^{1/2})median\{|\mathcal{W}_{1,k}|,k=1,.\dots,N\}/0.6745$ \citep{PercivalWalden2000}.
      
 Using the result of \cite{fanwang2008}, the jump variation is then estimated by the sum of the squares of all the estimated jump sizes:
 \begin{equation}
\widehat{JV}_{t,h}^W=\sum_{l=1}^{N_t} (\bar{y}_{t,h,\hat{\tau}_{l+}}-\bar{y}_{t,h,\hat{\tau}_{l-}})^2.
  \end{equation}
 thus we are able to estimate the jump variation from the process consistently with the convergence rate $N^{-1/4}$
\begin{equation} 
\widehat{JV}_{t,h}^W\overset{p}{\to}JV_{t,h}.
\end{equation}

In the following analysis, we will be able to separate the continuous part of the price process containing noise from the jump variation. This result can be found in \cite{fanwang2008} and it states that the jump-adjusted process $y_{t,h}^{(J)}=y_{t,h}-\widehat{JV}_{t,h}^W$ converges in probability to the continuous part without jumps, the integrated variance. Thus, if we are able to deal with the noise in $y_{t,h}^{(J)}$, we will be able to estimate the \textit{true} $IV_{t,h}$.

\subsection{Jump wavelet two scale realized variance estimator}
Finally, let us propose an estimator of realized variance that is able to estimate jumps from the process consistently and with $N\to\infty$, it is be able to recover the true integrated variance from noisy data. Moreover, we can use it to decompose the integrated variance into $J^m+1$ components. In the final estimator, we utilize what we already know: the TSRV estimator of \cite{zhang2005}, the wavelet-based realized variance estimator (Eq. \ref{wrv}) and the jump detection method proposed by previous section. Let $\widehat{RV}_{t,h}^{(estimator,J)}$ denote an estimator of realized variance over $\left[t-h,t\right]$, for $0 \le h \le t \le T$, on the jump-adjusted observed data, $y_{t,h}^{(J)}=y_{t,h}-\sum_{l=1}^{N_t} J_l$. The jump-adjusted wavelet two-scale realized variance estimator is defined as:
\begin{equation}
\label{jwtsrv}
\widehat{RV}_{t,h}^{(JWTSRV)}=\sum_{j=1}^{J^m+1}\widehat{RV}_{j,t,h}^{(JWTSRV)}=\sum_{j=1}^{J^m+1}\left(\widehat{RV}_{j,t,h}^{(W,J)}-\frac{\bar{N}}{N} \widehat{RV}_{j,t,h}^{(WRV,J)}\right),
\end{equation}
where $\widehat{RV}_{j,t,h}^{(W,J)}=\frac{1}{G} \sum_{g=1}^G  \sum_{k=1}^N \mathcal{W}_{j,t-h+{\frac{k}{N}h}}^2 $ obtained from wavelet coefficient estimates on a grid of size $\bar{N}=N/G$ and $\widehat{RV}_{j,t,h}^{(WRV,J)}=\sum_{k=1}^N \mathcal{W}_{j,t-h+{\frac{k}{N}h}}^2$ obtained from wavelet coefficient estimates on the jump-adjusted observed data, $y_{t,h}^{(J)}$.

The estimator JWTSRV in Eq.(\ref{jwtsrv}) uses jump-adjusted data $y_{t,h}^{(J)}=y_{t,h}-\widehat{JV}_{t,h}^W$ which are further decomposed by the wavelet transform (MODWT) to $j=1,...,J^m+1$ components. Final estimator is the sum of TSRV \citep{zhang2005} estimates on every particular component $j$. Since the TSRV estimator has rather slow rate of convergence of $N^{-1/6}$ and the wavelet MODWT estimator of realized variance has the rate of convergence $N^{-1/2}$, the speed of convergence of the JWTSRV components will be also $N^{-1/6}$. It is clear however, that the wavelet decomposition do not slow down the overall speed of convergence of the TSRV estimator compared to the JWTSRV, as well as it does not increase the asymptotic variance. Hence we can write:

\begin{equation} 
\widehat{RV}_{t,h}^{(JWTSRV)} \overset{p}{\to}IV_{t,h}
\end{equation}

Thus the JWTSRV is a consistent estimator of the integrated variance as it converges in probability to the \textit{true} integrated variance $IV_{t,h}$ of the process $p_t$. In the next section, we will test the finite sample properties and we show that variance of the JWTSRV is naturally inherited from TSRV.
%
%
In small samples, a small sample refinement can be constructed \citep{zhang2005}:

\begin{equation}
\label{smallsampleJWTSRV}
\widehat{RV}_{t,h}^{(JWTSRV,adj)}=\left(1-\frac{\bar{N}}{N} \right)^{(-1)} \widehat{RV}_{t,h}^{(JWTSRV)}.
\end{equation} 
When referring to the realized volatility estimated using our JWTSRV estimator, we will refer to the $\sqrt{\widehat{RV}_{t,h}^{(JWTSRV,adj)}}$.


\section{Numerical study of the small sample performance of the estimators \label{ch3}}

In this section, we study the small sample performance of all estimators using Monte Carlo simulations designed to capture the real nature of the data. We use several experiments using different volatility models, including a fractional stochastic volatility model capturing long memory in volatility and add numerical study of the forecasting performance of the estimators. Each experiment compares the performance of the realized variation estimator, the bipower variation estimator, the two-scale realized volatility, the realized kernel, and the jump wavelet two-scale realized variation defined by Eq. (\ref{jwtsrv}). All the estimators are adjusted for small sample bias, similarly to Eq. (\ref{smallsampleJWTSRV}). For convenience, we refer to the estimators in the description of the results as RV, BV, TSRV, RK and JWTSRV, respectively. Moreover, we also compare the minimum variance estimators TSRV$^*$ and JWTSRV$^*$, which minimize the total asymptotic variance of the estimators \citep{zhang2005}.

\subsection{Jump-diffusion model with stochastic volatility}
The first data generating model we assume in our study is a one-factor jump-diffusion model with stochastic volatility, described by the following equations: 
\begin{eqnarray}
\label{MCmodel1}
\nonumber dX_t&=&(\mu-\sigma_t^2/2)dt+\sigma_t dW_{x,t}+c_t dN_t \\
d\sigma_t^2&=&\kappa(\alpha-\sigma_t^2)dt+\gamma \sigma_t dW_{y,t},
\end{eqnarray}
where $W_x$ and $W_y$ are standard Brownian motions with correlation $\rho$, and $c_t dN_t$ is a compound Poisson process with random jump size distributed as $N\sim(0,\sigma_J)$. We set the parameters to values which are reasonable for a stock price, as in \cite{zhang2005}, who used model \ref{MCmodel1} without jumps, $\mu=0.05$, $\alpha=0.04$, $\kappa=5$, $\gamma=0.5$, $\rho=-0.5$ and $\sigma_J=0.025$. The volatility parameters satisfy Feller's condition $2\kappa\alpha\ge\gamma^2$, which keeps the volatility process away from the zero boundary. We generate $10,000$ independent sample paths\footnote{We have also computed the results for lower number of simulations, up to 1,000 generated independent sample paths and we found that the results do not change at all. These results are available upon request from authors.} of the process using the Euler scheme at a time interval of $\delta=1 s$, each with $6.5\times60\times60$ ($=23,400$) steps, corresponding to a 6.5 trading hour day. On each simulated path, we estimate $IV_{t,h}$ over $t=1$ day, as the parameter values are annualized (i.e., $t=1/252$). The results are computed for sampling of 5 minutes (78 observations) for RV, BV, TSRV, RK and JWTSRV, as well as for the optimal sampling frequency found by minimizing the total asymptotic variance of the estimators for TSRV$^*$ and JWTSRV$^*$.

We repeat the simulation with different levels of noise as well as different numbers of jumps. We assume that the market microstructure noise, $\epsilon_t$, comes from a Gaussian distribution with different standard deviations: $(E[\epsilon^2])^{1/2}=\left\{0,0.0005,0.001,0.0015\right\}$. Thus, the first simulated model, $(E[\epsilon^2])^{1/2}=0$, has zero noise. The remaining three models have levels of microstructure noise corresponding to 0.05\%, 0.1\% and 0.15\% of the value of the asset price.

\begin{table}[t]
\caption{Bias (variance in parenthesis) $\times 10^4$ of all estimators from 10,000 simulations of jump-diffusion model with $\epsilon_1=0$, $\epsilon_2=0.0005$, $\epsilon_3=0.001$, $\epsilon_4=0.0015$. RV -- 5 min. realized variance estimator, BV -- 5 min. bipower variation estimator, TSRV -- 5 min. two-scale realized volatility, JWTSRV -- 5 min. jump wavelet two-scale realized variance. TSRV$^*$ and JWTSRV$^*$ are minimum variance estimators, and RK is Realized Kernel.}
\centering
\tiny
\begin{tabular}{lrrrrrrr}
\toprule
& \multicolumn{1}{c}{\textbf{RV}}&\multicolumn{1}{c}{\textbf{BV}}&\multicolumn{1}{c}{\textbf{TSRV}}&\multicolumn{1}{c}{\textbf{TSRV$^*$}}&\multicolumn{1}{c}{\textbf{RK}} &\multicolumn{1}{c}{\textbf{JWTSRV}}&\multicolumn{1}{c}{\textbf{JWTSRV$^*$}} \\
\midrule
 \multicolumn{8}{c}{\textbf{No Jumps}} \\
$\epsilon_1$ &  0.90 (0.65) &  -4.13 (0.82) &  -6.03 (0.43) &  -0.28 (0.02) & -15.18 (2.51) &  -6.08 (0.43) &  -0.37 (0.02) \\ 
$\epsilon_2$ &100.10 (0.93) &  97.36 (1.18) &  -5.25 (0.45) &   0.98 (0.51) &  -4.40 (2.63) &  -3.86 (0.45) &   2.29 (0.52) \\ 
$\epsilon_3$ &394.14 (2.10) & 412.43 (2.87) &  -5.15 (0.45) &  -1.31 (0.90) &  19.66 (2.91) &   0.19 (0.48) &   3.95 (0.93) \\ 
$\epsilon_4$ &885.81 (5.40) & 949.39 (8.00) &  -4.52 (0.43) &  -0.47 (1.34) &  52.94 (3.13) &   7.71 (0.58) &  11.93 (1.48) \\ 
 \multicolumn{8}{c}{\textbf{One Jump}}\\ 
$\epsilon_1$ &247.73 (19.31) &  53.84 (1.85) & 236.63 (18.64) & 245.55 (18.09) & 225.41 (23.19) &  -5.64 (0.44) &  -0.25 (0.02) \\ 
$\epsilon_2$ &354.79 (20.91) & 164.24 (2.77) & 246.24 (19.67) & 253.69 (19.61) & 241.88 (23.10) &  -0.35 (0.48) &   4.36 (0.52) \\ 
$\epsilon_3$ &648.69 (23.12) & 495.58 (5.15) & 241.06 (19.79) & 251.24 (20.44) & 260.10 (25.62) &  18.12 (0.64) &  23.94 (1.10) \\ 
$\epsilon_4$ &1139.00 (27.54) & 1044.80 (10.79) & 248.00 (20.30) & 256.50 (21.02) & 303.39 (25.25) &  58.29 (1.41) &  64.39 (2.29) \\ 
 \multicolumn{8}{c}{\textbf{Two Jumps}}\\ 
$\epsilon_1$ &503.32 (41.12) & 117.87 (3.84) & 489.24 (39.47) & 501.61 (38.99) & 471.67 (47.36) &  -5.27 (0.43) &  -0.36 (0.02) \\ 
$\epsilon_2$ &616.80 (41.99) & 237.65 (4.56) & 500.37 (39.51) & 513.15 (39.69) & 489.82 (45.65) &   3.43 (0.49) &   7.41 (0.54) \\ 
$\epsilon_3$ &910.28 (44.71) & 582.94 (7.67) & 499.52 (39.83) & 508.95 (39.52) & 517.36 (48.27) &  38.99 (0.81) &  43.39 (1.25) \\ 
$\epsilon_4$ &1398.40 (47.55) & 1160.20 (15.04) & 496.34 (39.15) & 505.27 (38.93) & 551.50 (47.75) & 108.73 (2.34) & 113.95 (3.06) \\ 
\multicolumn{8}{c}{\textbf{Three Jumps}}\\ 
$\epsilon_1$ &772.53 (62.38) & 191.00 (6.58) & 753.28 (60.11) & 766.80 (58.86) & 730.70 (72.17) &  -5.62 (0.46) &  -0.37 (0.02) \\ 
$\epsilon_2$ &858.07 (61.60) & 312.01 (7.34) & 741.10 (58.62) & 759.90 (58.56) & 720.73 (68.89) &   6.04 (0.51) &  10.21 (0.53) \\ 
$\epsilon_3$ &1169.30 (68.71) & 671.31 (10.71) & 756.73 (61.89) & 767.36 (60.72) & 769.49 (74.86) &  59.15 (0.95) &  61.90 (1.37) \\ 
$\epsilon_4$ &1650.50 (69.52) & 1257.80 (18.55) & 742.31 (58.93) & 757.31 (59.37) & 787.06 (71.48) & 160.10 (3.19) & 167.24 (3.94) \\
\bottomrule
\label{model_jumpdiff}
\end{tabular}
\end{table}

Moreover, we add different amounts of jumps, controlled by intensity $\lambda$ from the Poisson process $c_t dN_t$. We start with $\lambda=0$, with model \ref{MCmodel1} reducing to a modification of the standard Heston volatility model without jumps, and continue with jump coefficients implying up to three jumps per day in the process. This number is realistic according to findings in the literature. The size of the jumps is controlled by parameter $\sigma_J$, which is set to 0.025, implying that a one standard deviation jump changes the price level by 2.5\%. Finally, we have 16 models with different levels of noise and numbers of jumps, and we compare the bias of all the estimators for each simulated day. 

Table \ref{model_jumpdiff} shows the results. The first model, without jumps, corresponds to the findings of \cite{zhang2005} and \cite{Sahalia2008}, although we add a higher level of noise to the simulations as suggested by the literature. The results show how robust the TSRV-based and RK estimators are to an increase in noise. Even a small increase in the magnitude of noise causes large bias in the other estimators, but the TSRV-based and RK estimators contain bias of order less than $10^{-4}$. What we add to the original results of \cite{zhang2005} and \cite{Sahalia2008} are jumps. While TSRV and RK are robust to an increase in noise, they are not robust to an increase in jumps. From the rest of the results, we can see how the wavelets detect all of the jumps in the process and the JWTSRV stays unbiased. From the results we can also see that with a mixture of relatively high noise and a large number of jumps in the process even the JWTSRV estimator suffers from bias. This suggests that jumps are sometimes indistinguishable from noise and remain undetected under the large noise. We can also see that the BV is able to deal with jumps to some extent, but is hurt heavily by noise.
 
\begin{table}[t]
\caption{Bias (variance in parenthesis) $\times 10^4$ of all estimators from 10,000 simulations of fractional stochastic volatility model with Hurst parameter $H=0.5$ with $\epsilon_1=0$, $\epsilon_2=0.0005$, $\epsilon_3=0.001$, $\epsilon_4=0.0015$. RV -- 5 min. realized variance estimator, BV -- 5 min. bipower variation estimator, TSRV -- 5 min. two-scale realized volatility, JWTSRV -- 5 min. jump wavelet two-scale realized variance. TSRV$^*$ and JWTSRV$^*$ are minimum variance estimators,  and RK is Realized Kernel.}
\tiny
\centering
\begin{tabular}{lrrrrrrr}
\toprule
& \multicolumn{1}{c}{\textbf{RV}}&\multicolumn{1}{c}{\textbf{BV}}&\multicolumn{1}{c}{\textbf{TSRV}}&\multicolumn{1}{c}{\textbf{TSRV$^*$}}&\multicolumn{1}{c}{\textbf{RK}} &\multicolumn{1}{c}{\textbf{JWTSRV}}&\multicolumn{1}{c}{\textbf{JWTSRV$^*$}} \\\midrule
 \multicolumn{8}{c}{\textbf{No Jumps}} \\
$\epsilon_1$ &  7.65 (10.51) & -19.57 (13.32) & -26.55 (6.86) &  -1.16 (0.23) & -66.40 (39.16) & -26.80 (6.86) &  -1.48 (0.23) \\ 
$\epsilon_2$ &104.62 (11.14) &  82.08 (14.41) & -26.79 (6.70) &  -0.41 (0.86) & -59.74 (38.38) & -25.51 (6.71) &   0.74 (0.86) \\ 
$\epsilon_3$ &407.48 (15.07) & 383.91 (19.41) & -23.47 (6.71) &  -0.98 (1.45) & -20.79 (41.69) & -18.27 (6.74) &   4.32 (1.48) \\ 
$\epsilon_4$ &896.20 (22.25) & 888.97 (29.63) & -25.28 (6.85) &  -5.32 (2.23) &  19.05 (44.65) & -13.75 (7.04) &   6.14 (2.36) \\ 
 \multicolumn{8}{c}{\textbf{One Jump}}\\ 
$\epsilon_1$ &254.70 (32.31) &  97.88 (18.00) & 219.71 (27.19) & 249.51 (19.81) & 167.85 (67.92) & -27.29 (6.69) &  -1.42 (0.25) \\ 
$\epsilon_2$ &356.65 (32.73) & 196.24 (19.36) & 219.05 (26.04) & 247.03 (18.80) & 184.96 (67.27) & -20.04 (6.68) &   4.05 (0.84) \\ 
$\epsilon_3$ &654.63 (37.40) & 507.79 (24.60) & 222.84 (27.44) & 249.36 (20.25) & 213.83 (70.42) &   1.88 (7.19) &  24.29 (1.64) \\ 
$\epsilon_4$ &1151.80 (45.69) & 1026.90 (36.71) & 226.66 (27.63) & 251.67 (21.35) & 266.50 (73.94) &  39.10 (8.15) &  60.25 (3.10) \\ 
 \multicolumn{8}{c}{\textbf{Two Jumps}}\\ 
$\epsilon_1$ &510.21 (53.31) & 217.50 (22.70) & 470.75 (47.16) & 505.47 (38.33) & 411.80 (97.74) & -25.56 (6.75) &  -0.42 (0.26) \\ 
$\epsilon_2$ &611.27 (57.48) & 317.64 (24.09) & 471.07 (49.70) & 506.63 (40.45) & 424.19 (101.42) & -20.62 (6.88) &   5.74 (0.87) \\ 
$\epsilon_3$ &914.79 (60.52) & 636.31 (30.92) & 476.78 (49.28) & 505.09 (40.70) & 466.31 (103.95) &  21.00 (7.38) &  42.32 (1.79) \\ 
$\epsilon_4$ &1396.70 (67.41) & 1155.20 (42.77) & 474.58 (47.27) & 504.16 (40.06) & 506.18 (103.54) &  93.30 (9.22) & 117.05 (4.00) \\ 
\multicolumn{8}{c}{\textbf{Three Jumps}}\\ 
$\epsilon_1$ &765.95 (78.40) & 346.13 (28.96) & 719.80 (69.95) & 750.18 (57.99) & 670.56 (134.88) & -23.75 (6.88) &  -1.96 (0.26) \\ 
$\epsilon_2$ &855.63 (76.82) & 436.22 (29.67) & 713.92 (66.84) & 750.49 (58.47) & 666.32 (127.53) & -15.91 (6.74) &   9.47 (0.88) \\ 
$\epsilon_3$ &1161.90 (81.72) & 762.38 (37.14) & 721.15 (68.21) & 758.76 (58.42) & 705.35 (134.87) &  35.08 (7.65) &  61.87 (1.96) \\ 
$\epsilon_4$ &1662.10 (95.44) & 1299.40 (52.60) & 722.50 (69.09) & 758.79 (59.70) & 746.30 (136.93) & 135.71 (10.19) & 162.66 (4.72) \\ 
\bottomrule
\label{model_h_05}
\end{tabular}
\end{table}

\subsection{Fractional stochastic volatility model}
Empirical evidence suggests that the volatility process may exhibit long memory. Previous models approximate this behavior, but a much more powerful class of models designed to capture long memory is known by the literature, namely, fractional Brownian motion. Instead of describing the solution and method of simulation of this class of models here, we rather point the interested reader to \cite{comte98} and \cite{marinucci99} for more details. 
 
\begin{table}[t]
\caption{Bias (variance in parenthesis) $\times 10^4$ of all estimators from 10,000 simulations of fractional stochastic volatility model with Hurst parameter $H=0.7$ with $\epsilon_1=0$, $\epsilon_2=0.0005$, $\epsilon_3=0.001$, $\epsilon_4=0.0015$. RV -- 5 min. realized variance estimator, BV -- 5 min. bipower variation estimator, TSRV -- 5 min. two-scale realized volatility, JWTSRV -- 5 min. jump wavelet two-scale realized variance. TSRV$^*$ and JWTSRV$^*$ are minimum variance estimators,  and RK is Realized Kernel.}
\tiny
\centering
\begin{tabular}{lrrrrrrr}
\toprule
& \multicolumn{1}{c}{\textbf{RV}}&\multicolumn{1}{c}{\textbf{BV}}&\multicolumn{1}{c}{\textbf{TSRV}}&\multicolumn{1}{c}{\textbf{TSRV$^*$}}&\multicolumn{1}{c}{\textbf{RK}} &\multicolumn{1}{c}{\textbf{JWTSRV}}&\multicolumn{1}{c}{\textbf{JWTSRV$^*$}} \\
\midrule
 \multicolumn{8}{c}{\textbf{No Jumps}} \\
$\epsilon_1$ &  9.47 (10.57) & -14.18 (13.44) & -25.81 (6.87) &  -0.62 (0.24) & -61.83 (39.91) & -26.17 (6.86) &  -0.94 (0.23) \\ 
$\epsilon_2$ &106.09 (11.24) &  78.59 (14.57) & -22.93 (6.73) &  -0.29 (0.84) & -49.16 (39.28) & -21.66 (6.75) &   0.86 (0.84) \\ 
$\epsilon_3$ &404.06 (14.44) & 380.66 (18.75) & -23.64 (6.79) &  -1.01 (1.45) & -13.44 (43.10) & -17.93 (6.88) &   4.50 (1.48) \\ 
$\epsilon_4$ &899.67 (22.67) & 895.53 (29.96) & -21.95 (6.89) &  -1.66 (2.19) &  32.94 (45.65) &  -9.40 (7.12) &  10.72 (2.33) \\ 
 \multicolumn{8}{c}{\textbf{One Jump}}\\ 
$\epsilon_1$ &260.24 (32.42) &  99.77 (17.58) & 226.07 (27.77) & 252.00 (19.93) & 175.24 (71.81) & -24.61 (6.75) &  -0.66 (0.24) \\ 
$\epsilon_2$ &361.23 (33.51) & 204.48 (19.56) & 222.55 (26.96) & 250.42 (19.87) & 194.31 (70.22) & -20.36 (6.68) &   3.47 (0.85) \\ 
$\epsilon_3$ &658.78 (36.67) & 507.47 (24.81) & 229.15 (26.77) & 253.33 (20.29) & 221.70 (71.31) &   1.16 (7.28) &  21.96 (1.62) \\ 
$\epsilon_4$ &1140.50 (47.95) & 1014.50 (37.09) & 221.27 (28.05) & 248.39 (22.10) & 260.55 (74.86) &  35.43 (8.07) &  61.22 (3.19) \\ 
 \multicolumn{8}{c}{\textbf{Two Jumps}}\\ 
$\epsilon_1$ &514.66 (55.01) & 219.27 (23.17) & 473.71 (48.22) & 503.64 (39.76) & 430.23 (100.78) & -23.00 (6.69) &  -1.45 (0.24) \\ 
$\epsilon_2$ &615.38 (57.57) & 318.85 (24.74) & 481.64 (49.01) & 508.26 (39.87) & 453.16 (102.60) & -14.61 (6.95) &   5.80 (0.87) \\ 
$\epsilon_3$ &903.32 (59.69) & 630.21 (30.51) & 470.80 (47.55) & 498.66 (39.14) & 467.10 (102.37) &  20.01 (7.23) &  41.69 (1.78) \\ 
$\epsilon_4$ &1400.90 (66.50) & 1164.00 (43.24) & 467.00 (46.26) & 505.48 (39.79) & 500.94 (102.72) &  86.73 (9.19) & 115.27 (4.02) \\ 
\multicolumn{8}{c}{\textbf{Three Jumps}}\\ 
$\epsilon_1$ &765.72 (78.57) & 340.76 (28.99) & 720.49 (70.78) & 754.51 (59.35) & 676.34 (135.14) & -28.45 (6.80) &  -1.85 (0.25) \\ 
$\epsilon_2$ &873.97 (79.58) & 452.01 (30.08) & 731.76 (70.61) & 765.04 (59.59) & 682.13 (134.89) & -12.12 (6.85) &  12.12 (0.88) \\ 
$\epsilon_3$ &1164.00 (82.53) & 767.45 (36.59) & 718.24 (67.72) & 752.01 (58.67) & 704.64 (132.81) &  38.63 (7.86) &  63.43 (1.96) \\ 
$\epsilon_4$ &1663.50 (91.73) & 1299.90 (48.60) & 731.80 (69.58) & 758.67 (59.03) & 756.10 (138.55) & 141.26 (10.54) & 161.96 (4.84) \\
\bottomrule
\label{model_h_07}
\end{tabular}
\end{table}

In our simulations, we use the fractional jump-diffusion model:
\begin{eqnarray}
\label{MCmodel2}
\nonumber dX_t=(\mu-\sigma_t^2/2)dt+\sigma_t dW_{x,t}+c_t dN_t \\
d\sigma_{H,t}^2=\kappa(\alpha-\sigma_{H,t}^2)dt+\gamma dW_{H,t},
\end{eqnarray}
where $W_x$ is a standard Brownian motion, $dW_{H,t}$ is a fractional Brownian motion (FBM) with Hurst parameter $H \in (0,1]$ and $c_t dN_t$ is a compound Poisson process with random jump size distributed as $N\sim(0,\sigma_J)$. We set the parameters to values $\mu=0.05$, $\alpha=0.2$, $\kappa=20$, $\gamma=0.012$ and $\sigma_J=0.025$ as in \cite{Sahalia2008}, although these authors use a process without jumps.

We generate $10,000$ independent sample paths\footnote{We have also computed the results for lower number of simulations, up to 1,000 generated independent sample paths and we found that the results do not change at all. These results are available upon request from authors.}  of the process using the Euler scheme at a time interval of $\delta=1 s$, each with $6.5\times60\times60$  ($=23,400$) steps, corresponding to 6.5 trading hours. The results are computed for sampling of 5 minutes (78 observations) for RV, BV, TSRV, RK and JWTSRV, as well as for the optimal sampling frequency found by minimizing the total asymptotic variance for TSRV$^*$ and JWTSRV$^*$. We again repeat the simulation with different levels of noise as well as different numbers of jumps. We assume that the market microstructure noise, $\epsilon_t$, comes from a Gaussian distribution with different standard deviations: $(E[\epsilon^2])^{1/2}=\left\{0,0.0005,0.001,0.0015\right\}$, and we again start without jumps, and continue with jump coefficients implying up to three jumps per day in the process. Finally, we have 16 models with different levels of noise and numbers of jumps, and we compare the bias of all the estimators for each simulated day on three processes with different long memory parameters.

\begin{table}[t]
\caption{Bias (variance in parenthesis) $\times 10^4$ of all estimators from 10,000 simulations of fractional stochastic volatility model with Hurst parameter $H=0.9$ with $\epsilon_1=0$, $\epsilon_2=0.0005$, $\epsilon_3=0.001$, $\epsilon_4=0.0015$. RV -- 5 min. realized variance estimator, BV -- 5 min. bipower variation estimator, TSRV -- 5 min. two-scale realized volatility, JWTSRV -- 5 min. jump wavelet two-scale realized variance. TSRV$^*$ and JWTSRV$^*$ are minimum variance estimators, and RK is Realized Kernel.}
\tiny
\centering
\begin{tabular}{lrrrrrrr}
\toprule
& \multicolumn{1}{c}{\textbf{RV}}&\multicolumn{1}{c}{\textbf{BV}}&\multicolumn{1}{c}{\textbf{TSRV}}&\multicolumn{1}{c}{\textbf{TSRV$^*$}}&\multicolumn{1}{c}{\textbf{RK}} &\multicolumn{1}{c}{\textbf{JWTSRV}}&\multicolumn{1}{c}{\textbf{JWTSRV$^*$}} \\
\midrule
 \multicolumn{8}{c}{\textbf{No Jumps}} \\
$\epsilon_1$ &  5.90 (10.50) & -19.58 (13.51) & -26.67 (6.78) &  -0.50 (0.25) & -61.72 (40.10) & -27.11 (6.78) &  -0.92 (0.25) \\ 
$\epsilon_2$ &110.47 (11.65) &  84.15 (14.78) & -22.39 (7.05) &   0.18 (0.83) & -47.77 (40.61) & -21.14 (7.07) &   1.33 (0.84) \\ 
$\epsilon_3$ &399.57 (15.33) & 372.67 (19.88) & -29.78 (6.77) &  -1.92 (1.45) & -34.79 (42.18) & -25.00 (6.83) &   3.31 (1.47) \\ 
$\epsilon_4$ &882.50 (22.98) & 879.81 (30.30) & -28.32 (6.74) &  -0.72 (2.18) &  14.36 (44.32) & -17.21 (6.93) &  10.63 (2.30) \\ 
 \multicolumn{8}{c}{\textbf{One Jump}}\\ 
$\epsilon_1$ &269.61 (35.26) & 100.30 (17.80) & 233.23 (29.92) & 258.49 (21.56) & 184.42 (73.31) & -25.58 (6.86) &  -2.19 (0.25) \\ 
$\epsilon_2$ &364.35 (34.40) & 200.35 (19.23) & 226.94 (28.28) & 258.79 (21.54) & 201.67 (71.72) & -21.96 (6.82) &   4.05 (0.85) \\ 
$\epsilon_3$ &648.93 (38.20) & 498.06 (24.94) & 218.29 (27.72) & 249.82 (20.78) & 214.87 (74.11) &  -5.09 (7.19) &  23.16 (1.66) \\ 
$\epsilon_4$ &1143.10 (44.73) & 1017.00 (35.55) & 221.50 (27.14) & 250.37 (21.52) & 255.73 (71.84) &  36.01 (8.13) &  60.87 (3.15) \\ 
 \multicolumn{8}{c}{\textbf{Two Jumps}}\\ 
$\epsilon_1$ &507.64 (54.73) & 217.73 (23.69) & 468.53 (48.86) & 499.98 (37.75) & 422.53 (106.44) & -27.23 (7.05) &  -1.50 (0.25) \\ 
$\epsilon_2$ &618.08 (57.83) & 323.80 (24.72) & 475.53 (49.28) & 505.72 (39.53) & 446.02 (102.66) & -13.93 (6.99) &   6.57 (0.88) \\ 
$\epsilon_3$ &902.48 (63.40) & 620.44 (30.54) & 470.85 (50.44) & 502.56 (40.29) & 462.64 (106.41) &  15.21 (7.49) &  43.52 (1.81) \\ 
$\epsilon_4$ &1399.10 (70.64) & 1156.50 (43.00) & 470.35 (49.76) & 498.97 (40.92) & 504.29 (109.07) &  87.73 (9.16) & 114.08 (3.94) \\ 
\multicolumn{8}{c}{\textbf{Three Jumps}}\\ 
$\epsilon_1$ &767.20 (77.56) & 337.59 (28.64) & 721.80 (68.54) & 755.62 (56.93) & 674.80 (130.51) & -25.42 (6.82) &  -2.66 (0.25) \\ 
$\epsilon_2$ &866.12 (78.84) & 443.31 (30.34) & 720.71 (69.21) & 754.90 (58.38) & 689.69 (134.96) & -13.72 (6.83) &  11.64 (0.91) \\ 
$\epsilon_3$ &1164.80 (83.67) & 759.78 (36.66) & 730.27 (69.86) & 758.64 (59.37) & 713.23 (135.13) &  41.37 (7.85) &  60.48 (1.93) \\ 
$\epsilon_4$ &1661.80 (93.00) & 1303.50 (50.63) & 724.24 (69.10) & 752.55 (59.56) & 762.91 (145.29) & 142.24 (10.54) & 163.84 (4.82) \\ 
\bottomrule
\label{model_h_09}
\end{tabular}
\end{table}

Increments of the volatility process with $H\in(0.5,1]$ exhibit the desired long memory. Thus we will study this model for a Hurst exponent equal to $H=\{0.5,0.7,0.9\}$. While the first case has independent increments, the second and third cases exhibit quite strong long memory processes in volatility.

Tables \ref{model_h_05}, \ref{model_h_07} and \ref{model_h_09} summarize the results for the different $H=\{0.5,0.7,0.9\}$, respectively. The results confirm exactly the same behavior for all the estimators as in the previous case without long memory. Thus we can conclude that our JWTSRV estimator is robust to jumps and noise on small samples even if we consider the volatility process with long memory, and it proved to be the best estimator of $IV_{t,h}$ even on small samples. While we studied only the in-sample performance of the estimator, we present the out-of-sample, or forecasting, performance in the next section.

\subsection{One-day-ahead forecasts of IV using JWTSRV}

One of the many potential useful applications of the proposed framework is volatility forecasting. In particular, the one-day-ahead return variation forecast, $var(p_{t+1}| \mathcal{F}_t)$, is of huge interest for practitioners. Thus we would like to study the forecasting ability of the proposed methodology as well. While we showed that the in-sample performance of the estimators is the same for different models and that the JWTSRV estimator tends to consistently estimate $IV_{t,h}$ regardless of the level of noise and number of jumps in the process, we will reduce our simulation scheme to model in Eq.(\ref{MCmodel1}) with a fixed level of noise and number of jumps. This setting will allow us to study the impact of noise and jumps on the forecasting performance of the estimators and to see if the JWTSRV holds its power and is able to forecast $var(p_{t+1}| \mathcal{F}_t)$. 

In this exercise, we follow the framework of \cite{Sahalia2008} closely. Denoting the annualized one day time interval $T_1-T_0=T_2-T_1$, 
\begin{equation}
\label{for1}
E\left[\sigma_{T_1}^2 | \mathcal{F}_{T_0}\right ]=e^{-\kappa(T_1-T_0)} \sigma_{T_0}^2+\alpha(1-e^{-\kappa(T_1-T_0)}),
\end{equation}
where $\sigma_t^2$ follows model in Eq. (\ref{MCmodel1}) and $\mathcal{F}_{T}=\sigma \{ \sigma_t^2;t\le T \}$ is the information set generated by the instantaneous variance process up to time $T$. If we use integration operators, we have
\begin{equation}
\label{for2}
E\left[\int_{T_0}^{T_1} \sigma_{t}^2 dt | \mathcal{F}_{T_0}\right]=\frac{1}{\kappa}(1-e^{-\kappa(T_1-T_0)})\sigma_{T_0}^2+\alpha(T_1-T_0)-\frac{\alpha}{\kappa}(1-e^{-\kappa(T_1-T_0)}).
\end{equation}
If we want to express the one-day-ahead forecast, we simply use Eq. (\ref{for1}) and Eq. (\ref{for2}) and we get:
\begin{equation}
\label{for3}
E\left[\int_{T_m}^{T_{m+1}} \sigma_{t}^2 dt | \mathcal{F}_{T_{m-1}}\right]=e^{-\kappa D} E\left[\int_{T_{m-1}}^{T_{m}} \sigma_{t}^2 dt | \mathcal{F}_{T_{m-1}}\right]+\alpha(1-e^{-\kappa D})D,
\end{equation}
where $D=T_{m+1}-T_m=T_m-T_{m-1}$. Eq. (\ref{for3}) is the exact conditional forecast of $\int_{T_m}^{T_{m+1}} \sigma_t^2 dt$, but it is not feasible, as $E\left[\int_{T_{m-1}}^{T_m} \sigma_{t}^2 dt | \mathcal{F}_{T_{m-1}}\right]$  is not observed in practice. But if we replace this term by the estimate of the integrated variance on day $m$ we arrive at a simple method for forecasting the integrated variance on day $m+1$. In empirical applications the true underlying model parameters are unknown and the properties of the observed data differ from the simulated ones, even though the simulations are based on estimated parameters on real-world data. Hence, the estimation is required to be realistic, and the AR(1) process seems to serve well in this case.

We use the simulation scheme for model in Eq.(\ref{MCmodel1}) from the previous section. This time, we simulate 101 ``continuous" sample paths over days $[0,T_1],\dots,[T_{99},T_{100}],$ $[T_{100},T_{101}]$, that is, $101\times23,400$ log returns. We split each simulated path into two parts. The first part, of $100\times23,400$, is used to estimate the time series of 100 daily integrated variations using the tested estimators. Then, the AR(1) model is used to estimate the coefficients of forecast Eq.(\ref{for3}), where the conditional expectation in the right-hand side is replaced by the estimated integrated variation. The second part, the last (101th) day, is saved for out-of-sample comparison purposes as the true integrated variance of the day, which is compared with the AR(1) forecast of the integrated variance for the $m+1$th day. This procedure is repeated for each simulated sample path of $101\times23,400$ log returns and all the estimators tested in the previous exercise.

We employ the traditional \cite{mincerzarnowitz} approach to assess the forecasting performance of the individual estimators and we compare alternative variance forecasts by projecting the true integrated variance on day $m+1$, $\int_{T_m}^{T_{m+1}} \sigma_{t}^2 dt$, on a constant and various estimator forecasts. For example, we evaluate the JWTSRV forecasting performance by running the following regression:
\begin{equation}
\label{mincerex}
IV_{T_m{+1}}=\alpha+\beta V_{T_{m+1|T_m}}^{JWTSRV}+\epsilon,
\end{equation}
where $V_{T_{m+1|T_m}}^{JWTSRV}$ is the one-day-ahead forecast of integrated variance from day $m$ to day $m+1$ using the AR(1) prediction. Thus, Eq.(\ref{mincerex}) regresses the true realized variance $IV_{T_{m+1}}$ from day $m+1$ on a constant and the variance forecast using the JWTSRV estimator. If the JWTSRV estimator performs well, the forecast should be unbiased and the forecast error is small. In other words, $\alpha=0$ and $\beta=1$, and the $R^2$ of the regression is close to 1. Thus we will test the null hypothesis of $H_0:\alpha=0$ and $H_0:\beta=1$ against the alternatives $H_A:\alpha \ne 0$ and $H_A:\beta \ne1$.

\begin{table}[t]
\caption{Out-of-sample Mincer-Zarnowitz regressions (Eq. \ref{mincer}) on model with no jumps. Results significant at 95\% are in bold; OLS standard errors in parenthesis.}

\tiny
\centering
\ra{1}
\begin{tabular}{lrrrrrrr}
\toprule
\multicolumn{7}{l}{Joint Mincer-Zarnowitz regression} \\
\midrule
&  \multicolumn{1}{c}{\textbf{const.}} & \multicolumn{1}{c}{\textbf{RV}}&\multicolumn{1}{c}{\textbf{BV}}&\multicolumn{1}{c}{\textbf{TSRV}}&\multicolumn{1}{c}{\textbf{RK}}&\multicolumn{1}{c}{\textbf{JWTSRV}}&\multicolumn{1}{c}{\textbf{$R^2$}} \\
\cmidrule{2-8}
& \textbf{-0.055} (0.003) & \textbf{1.568} (0.099) & \textbf{-0.440} (0.103) & &				&				&  0.895 \\
& \textbf{0.009} (0.003) & 0.072 (0.093) & \textbf{-0.172 (0.078)} & \textbf{1.070} (0.039)  & & 				& 0.941 \\
 & \textbf{0.009} (0.003) & 0.082 (0.090) & -0.108 (0.077) & \textbf{1.184} (0.0407) & \textbf{-0.199} (0.027) & & 0.944 \\
  & \textbf{0.009} (0.003) & 0.082 (0.090) & -0.109 (0.077) & \textbf{1.054} (0.293) & \textbf{-0.199} (0.027)  & 0.131 (0.294) &0.944 \\
\midrule
\multicolumn{7}{l}{Individual Mincer-Zarnowitz regression} \\
\midrule
&  \multicolumn{1}{c}{\textbf{const.}} & \multicolumn{1}{c}{\textbf{RV}}&\multicolumn{1}{c}{\textbf{BV}}&\multicolumn{1}{c}{\textbf{TSRV}}&\multicolumn{1}{c}{\textbf{RK}}&\multicolumn{1}{c}{\textbf{JWTSRV}}&\multicolumn{1}{c}{\textbf{$R^2$}} \\
\cmidrule{2-8}
RV & \textbf{-0.059} (0.003) & \textbf{1.145} (0.013) &		&		&		&		& 0.893 \\
BV &  \textbf{-0.063} (0.003)	& 				      & \textbf{1.167} (0.014) & & & & 0.869 \\
TSRV & -0.002 (0.002)		& 					&		& \textbf{0.995} (0.008)	& & & 0.940 \\
RK & -0.002 (0.003)		& 					&		& 	& \textbf{1.017} (0.016) & & 0.807 \\
JWTSRV & 0.001 (0.002) & 			 	& &	& & \textbf{0.997} (0.008)	& 0.939 \\
\midrule
\multicolumn{6}{l}{Mincer-Zarnowitz regression for minimum variance TSRV estimators} \\
\midrule
&  \multicolumn{1}{c}{\textbf{const.}} & & &\multicolumn{1}{c}{\textbf{TSRV$^*$}}& &\multicolumn{1}{c}{\textbf{JWTSRV$^*$}}&\multicolumn{1}{c}{\textbf{$R^2$}} \\
\cmidrule{2-8}
TSRV$^*$ & -.002 (0.001)  &						&		& \textbf{0.993} (0.006) & & & 0.959 \\
JWTSRV$^*$ & 0.001 (0.001) &					&		&				&	& \textbf{0.996} (0.006) &    0.959 \\
 \bottomrule
\end{tabular}
\label{MincerJ0}
\end{table}

In our simulations, we study a Mincer-Zarnowitz style regression combining several estimators:
\begin{eqnarray}
\label{mincer}
\nonumber 
\left\{ IV_{T_{m+1}} \right\}_j & = & \alpha+\beta_1 \left\{ V_{T_{m+1|T_m}}^{RV} \right\}_j+\beta_2 \left\{ V_{T_{m+1|T_m}}^{BV} \right\}_j+\beta_3 \left\{ V_{T_{m+1|T_m}}^{TSRV} \right\}_j \\
& & + \beta_4 \left\{ V_{T_{m+1|T_m}}^{RK} \right\}_j+\beta_5 \left\{ V_{T_{m+1|T_m}}^{JWTSRV} \right\}_j +\epsilon_j
\end{eqnarray}
for $j=1,\dots,10,000$ simulated sample paths. $V_{T_{m+1|T_m}}^{\mathcal{M}}$ is the one-day-ahead forecast of integrated variance from day $m$ to day $m+1$ given by the AR(1) model for the time series of daily variance estimated by the $\mathcal{M}$ estimator of realized variance. Eq. (\ref{mincer}) can be naturally interpreted as a variance forecast encompassing regression, as a coefficient significantly different from zero implies that the information in that particular forecast is not included in the forecasts of other models. To test the robustness of the results, we also run individual regressions where we consider only a constant and a single forecasting model. Thus we run four separate regressions to supplement the joint regression from Eq.(\ref{mincer}). 

\begin{table}[t]
\caption{Out-of-sample Mincer-Zarnowitz regressions (Eq. \ref{mincer}) on model with 1 jump. Results significant at 95\% are in bold; OLS standard errors in parenthesis.}
\tiny
\centering
\ra{1}
\begin{tabular}{lrrrrrrr}
\toprule
\multicolumn{7}{l}{Joint Mincer-Zarnowitz regression} \\
\midrule
&  \multicolumn{1}{c}{\textbf{const.}} & \multicolumn{1}{c}{\textbf{RV}}&\multicolumn{1}{c}{\textbf{BV}}&\multicolumn{1}{c}{\textbf{TSRV}}&\multicolumn{1}{c}{\textbf{RK}}&\multicolumn{1}{c}{\textbf{JWTSRV}}&\multicolumn{1}{c}{\textbf{$R^2$}} \\
\cmidrule{2-8}
& \textbf{-0.032} (0.006) & \textbf{-0.500} (0.045) & \textbf{1.538} (0.037) & & & & 0.811 \\
& \textbf{0.032} (0.010) & \textbf{-1.857} (0.185) & \textbf{1.512} (0.036) & \textbf{1.251} (0.166) & & & 0.822 \\
& \textbf{0.032} (0.010) & \textbf{-1.873} (0.186) & \textbf{1.514} (0.036) & \textbf{1.182} (0.183) & 0.088 (0.098) & & 0.822 \\
& 0.000 (0.006) & 0.129 (0.122) & -0.045 (0.043) & -0.042 (0.113) & \textbf{-0.204} (0.059) & \textbf{1.078} (0.026) & 0.936 \\
\midrule
\multicolumn{7}{l}{Individual Mincer-Zarnowitz regression} \\
\midrule
&  \multicolumn{1}{c}{\textbf{const.}} & \multicolumn{1}{c}{\textbf{RV}}&\multicolumn{1}{c}{\textbf{BV}}&\multicolumn{1}{c}{\textbf{TSRV}}&\multicolumn{1}{c}{\textbf{RK}}&\multicolumn{1}{c}{\textbf{JWTSRV}}&\multicolumn{1}{c}{\textbf{$R^2$}} \\
\cmidrule{2-8} 
RV & \textbf{-0.100} (0.009) & \textbf{1.123} (0.037) & & & & & 0.480 \\
BV &  \textbf{-0.079} (0.005) & & \textbf{1.181} (0.019) & & & & 0.788 \\
TSRV & \textbf{-0.050} (0.007) & & & \textbf{1.028} (0.032) & & & 0.500 \\
RK & \textbf{-0.051} (0.008) & & & & \textbf{1.041} (0.035) & & 0.476 \\
JWTSRV & -0.003 (0.002) & & & & & \textbf{1.024} (0.009) & 0.935 \\
\midrule
\multicolumn{7}{l}{Mincer-Zarnowitz regression for minimum variance TSRV estimators} \\
\midrule
&  \multicolumn{1}{c}{\textbf{const.}} & & &\multicolumn{1}{c}{\textbf{TSRV$^*$}}& &\multicolumn{1}{c}{\textbf{JWTSRV$^*$}}&\multicolumn{1}{c}{\textbf{$R^2$}} \\

\cmidrule{2-8}
TSRV$^*$ & \textbf{-0.049} (0.007) & & &\textbf{1.021} (0.032) & & & 0.506 \\
JWTSRV$^*$ &  -0.003 (0.002) & & & & & \textbf{1.014} (0.007) & 0.957 \\
 \bottomrule
\end{tabular}
\label{MincerJ1}
\end{table}

\subsubsection{Forecasting without jumps}
We run the simulations for two model settings using model in Eq.(\ref{MCmodel1}) with one jump and with no jumps. Let us start with the model without jumps first. The OLS estimates of all the forecast evaluation regressions for the model without jumps are reported in Table \ref{MincerJ0}. The results suggests that the TSRV performs as the best forecasting vehicle. Comparing the individual regressions, the TSRV has the highest $R^2$ and the coefficient closest to 1 with an insignificant coefficient, which suggests that the forecasts of the TSRV are biased only very slightly (as the coefficient is significantly different from 1). When looking at the joint regressions, we can see that the addition of all the other estimators does not improve this result. Moreover, when the TSRV is included in the regression, it is the only significant estimator, meaning that none of the other estimators has additional information not included in the TSRV forecast. In other words, adding the other estimators' forecasts to the TSRV brings no additional explanatory power to the regression. The JWTSRV forecast has the same performance as the simple TSRV, as there are no jumps in the simulated process, thus the asymptotic behavior of these two estimators should be the same. The JWTSRV is expected to have much better performance in the simulations where we include jumps. All the estimators are estimated with a 5-minute sampling frequency.

In addition, we provide results for the optimal sampling minimizing variance of the estimator in the last part of the table. The TSRV$^*$ with optimally chosen sampling outperforms the 5 min. TSRV. The JWTSRV$^*$ again has the same performance as expected.

\subsubsection{Forecasting with jumps}
Let's see how the results change when we add a single jump to the simulated model. The OLS estimates of all the forecast evaluation regressions for the model with jumps are reported in Table \ref{MincerJ1}. Looking at the results of the individual regressions, one can see that the JWTSRV largely outperforms all the other estimators, with $R^2$ close to the results from the model without jumps from the previous section. This suggests that the JWTSRV is robust to jumps even when we consider forecasting. The joint regression confirms this result. The regression including all the forecasts using the four considered estimators has the largest explanatory power. Moreover, the coefficient of the JWTSRV is significant, while the other coefficients are not significant, suggesting that the other estimators carry no additional information. Taking the JWTSRV forecasts away from the regression results in much lower $R^2$. It is interesting to note that in this case all the other coefficients are significant, suggesting multicollinearity caused by jumps in the process. The reader can also note how the addition of the BV improves the result. In fact, the BV rules the TSRV, with much higher $R^2$. In fact, the BV is used for jump detection, so this finding confirms the results from the literature.

In addition, we include results for optimal sampling, which minimizes the variance of the TSRV-based estimators. In this case, we can see that the result improves and the JWTSRV$^*$ yields the best result. 

To conclude this section, the results suggest that when the JWTSRV estimator is used for variance forecasting in the presence of jumps and noise, the forecasts will be unbiased. This makes the JWTSRV estimator a very powerful tool for forecasting the variance of stock market returns. With the theoretical results in hand, we can move to empirical examples and use the JWTSRV to forecast the volatility of real-world data.


\section{Decomposition of empirical volatility}

In this section, we turn our focus to real-world data estimation of the proposed theory. We will test several integrated volatility estimators in comparison to our JWTSRV estimator and study their distributional properties. The JWTSRV proved to have lowest bias in the Monte Carlo simulations, thus we also expect it to have the best performance on the real data set. 

\subsection{Data description}
Foreign exchange future contracts are traded on the Chicago Mercantile Exchange (CME) on a 24-hour basis. As these markets are among the most liquid, they are suitable for analysis of high-frequency data. We will estimate the realized volatility of British pound (GBP), Swiss franc (CHF) and euro (EUR) futures. All contracts are quoted in the unit value of the foreign currency in US dollars. It is advantageous to use currency futures data for the analysis instead of spot currency prices, as they embed interest rate differentials and do not suffer from additional microstructure noise coming from over-the-counter trading. The cleaned data are available from Tick Data, Inc.\footnote{http://www.tickdata.com/}

It is very important to look first at the changes in the trading system before we proceed with the estimation on the data. In August 2003, for example, the CME launched the Globex trading platform, and for the first time ever in a single month, the trading volume on the electronic trading platform exceeded 1 million contracts every day. On Monday, December 18, 2006, the CME Globex(R) electronic trading platform started offering nearly continuous trading. More precisely, the trading cycle became 23 hours a day (from 5:00 pm on the previous day until 4:00 pm on current day, with a one-hour break in continuous trading), from 5:00 pm on Sunday until 4:00 pm on Friday. These changes certainly had a dramatic impact on trading activity and the amount of information available, resulting in difficulties in comparing the estimators on the pre-2003 data, the 2003--2006 data and the post--2006 data. For this reason, we restrict our analysis to a sample period extending from January 5, 2007 through November 17, 2010, which contains the most recent financial crisis. The futures contracts we use are automatically rolled over to provide continuous price records, so we do not have to deal with different maturities.

The tick-by-tick transactions are recorded in Chicago Time, referred to as Central Standard Time (CST). Therefore, in a given day, trading activity starts at 5:00 pm CST in Asia, continues in Europe followed by North America, and finally closes at 4:00 pm in Australia. To exclude potential jumps due to the one-hour gap in trading, we redefine the day in accordance with the electronic trading system. Moreover, we eliminate transactions executed on Saturdays and Sundays, US federal holidays, December 24 to 26, and December 31 to January 2, because of the low activity on these days, which could lead to estimation bias. Finally, we are left with 944 days in the sample. Looking more deeply at higher frequencies, we find a large amount of multiple transactions happening exactly at the same time stamp. We use the arithmetic average for all observations with the same time stamp.

\begin{table}
\scriptsize
\caption{The table summarizes the  daily log-return distributions of GBP, CHF and EUR futures. The sample period extends from January 5, 2007 through November 17, 2010, accounting for a total of 944 observations.}
\centering
\ra{1.2}
\begin{tabular}{lrrrr}
\toprule
& Mean & St.dev. & Skew. & Kurt.\\ 
\midrule
\textbf{GBP}  &    0.0001  &  0.0119  & -0.3852 &   4.4356 \\
\textbf{CHF}  &     0.0002  &  0.0068  &  0.2440  &  5.4662 \\
\textbf{EUR}  & 0.0002  &  0.0099   & 0.1536  &  4.4951 \\
\bottomrule
\end{tabular}
\label{summarystats}
\end{table}

\subsection{Statistical properties of unconditional return and integrated volatility}
Having prepared the data, we can estimate the integrated volatilities and study their statistical properties as well as the properties of the daily unconditional returns. For each futures contract, the daily integrated volatility is estimated using the square root of realized variance estimator, the bipower variation estimator, the two-scale realized volatility, the realized kernel and the jump wavelet two-scale realized variance defined by Eq.(\ref{jwtsrv}). All the estimators are adjusted for small sample bias. For convenience, we refer to the estimators in the description of the results as RV, BV, TSRV, RK and JWTSRV, respectively. The RV and BV estimates are estimated on 5-min log-returns. The TSRV and the JWTSRV are estimated using a slow time scale of 5 minutes.

\begin{table}
\scriptsize
\caption{The table summarizes the  daily standardized daily log-return distributions for GBP, CHF and EUR futures using $r_t/IV_t^{1/2}$ and daily distributions of integrated volatility $IV_t^{1/2}$. Integrated volatility $IV_t^{1/2}$ is estimated using the RV, the BV on 5-min. log-returns, and the TSRV and JWTSRV on 5 minutes for a slow time scale and the RK. The sample period extends from January 5, 2007 through November 17, 2010, accounting for a total of 944 observations.}
\centering
\ra{1.2}
\begin{tabular}{lrrrrlrrrr}
\toprule
 & \multicolumn{4}{c}{\textbf{Distributions of $r_t/IV_t^{1/2}$}} & &  \multicolumn{4}{c}{\textbf{Distributions of $IV_t^{1/2}$}}\\
 \cmidrule{2-5}  \cmidrule{7-10}
 & \multicolumn{4}{c}{\textbf{GBP futures}} & & \multicolumn{4}{c}{\textbf{GBP futures}} \\
 \cmidrule{2-5}  \cmidrule{7-10}
 & Mean & St.dev. & Skew. & Kurt.&  & Mean & St.dev. & Skew. & Kurt.\\ 
 \cmidrule{2-5}  \cmidrule{7-10}
  RV &  0.0419  &  0.8834 &   -0.0880  &  2.6029 &  RV&    0.0075  &  0.0038  &  1.8394   & 7.5736 \\
  BV & 0.0448   & 0.9266 &  -0.0669  &  2.6941  & BV &     0.0073  &  0.0037  &  1.7336  &  6.7996\\
  TSRV &  0.0451   & 0.9026 &  -0.0710 &   2.5744 &   TSRV &     0.0073  &  0.0037  &  1.7611  &  7.0767\\
   RK & 0.0458 &   0.9406 &  -0.0757 &   2.5162 & RK & 0.0070   & 0.0037  &  1.8201 &   7.6473 \\
   JWTSRV & 0.0489   & 0.9035 &  -0.0710  &  2.7512 &    JWTSRV &   0.0071  &  0.0037  &  1.7629  &  7.0112\\
 \cmidrule{2-5}  \cmidrule{7-10}
 & \multicolumn{4}{c}{\textbf{CHF futures}} & & \multicolumn{4}{c}{\textbf{CHF futures}} \\
 \cmidrule{2-5}  \cmidrule{7-10}
      RV &  0.0238  &  0.8959 &   0.0380  &  2.6272 &   RV &    0.0076  &  0.0029  &  1.6875  &  8.2794\\
  BV & 0.0272  &  0.9424 &   0.0727  &  2.7020 &   BV &     0.0073  &  0.0028 &   1.5696  &  7.5983\\
  TSRV & 0.0278  &  0.9180 &   0.0568  &  2.6161 &   TSRV &     0.0073  &  0.0028  &  1.5572  &  7.3379\\
  RK & 0.0281  &  0.9530  &  0.0425  &  2.5371 & RK & 0.0070  &  0.0028  &  1.8179  &  9.9149 \\
  JWTSRV & 0.0389  &  0.9253 &   0.0611 &   2.7170 &   JWTSRV &     0.0070  &  0.0026  &  1.4359  &  6.5452\\
 \cmidrule{2-5}  \cmidrule{7-10}
 & \multicolumn{4}{c}{\textbf{EUR futures}} & & \multicolumn{4}{c}{\textbf{EUR futures}} \\
 \cmidrule{2-5}  \cmidrule{7-10}
   RV &  0.0379 &   0.9550  & -0.0215  &  2.5728 &   RV &       0.0068 &   0.0031  &  1.4785  &  5.8493\\
  BV & 0.0410  &  0.9970  & -0.0271  &  2.6219 &   BV &       0.0066  &  0.0031  &  1.5001  &  5.9803\\
  TSRV & 0.0397  &  0.9638  & -0.0133  &  2.5502 &   TSRV &       0.0068  &  0.0031  &  1.4263  &  5.4871\\
  RK & 0.0415  &  0.9898 &  -0.0069  &  2.4497 & RK & 0.0065  &  0.0031  &  1.5351  &  6.2713 \\
  JWTSRV &0.0452 &   0.9587 &   0.0014  &  2.8144 &   JWTSRV &       0.0064 &   0.0030  &  1.4345  &  5.4716\\
\bottomrule
\end{tabular}
\label{summarystats2}
\end{table}

Table \ref{summarystats} presents the summary statistics for the daily log-returns of GBP, CHF and EUR futures over the sample period, $t=1,\dots,944$, i.e., January 5, 2007 to November 17, 2010. The summary statistics display an average return very close to zero, skewness, and excess kurtosis which is consistent with the large empirical literature started probably by \cite{fama} and \cite{mandelbrot}. As observed by \cite{abde2001}, when the log-returns are standardized by the integrated volatility, $r_t/IV_t^{1/2}$, the unconditional returns are very close to a Gaussian distribution. 

\begin{figure}
   \centering
   \includegraphics[width=4in]{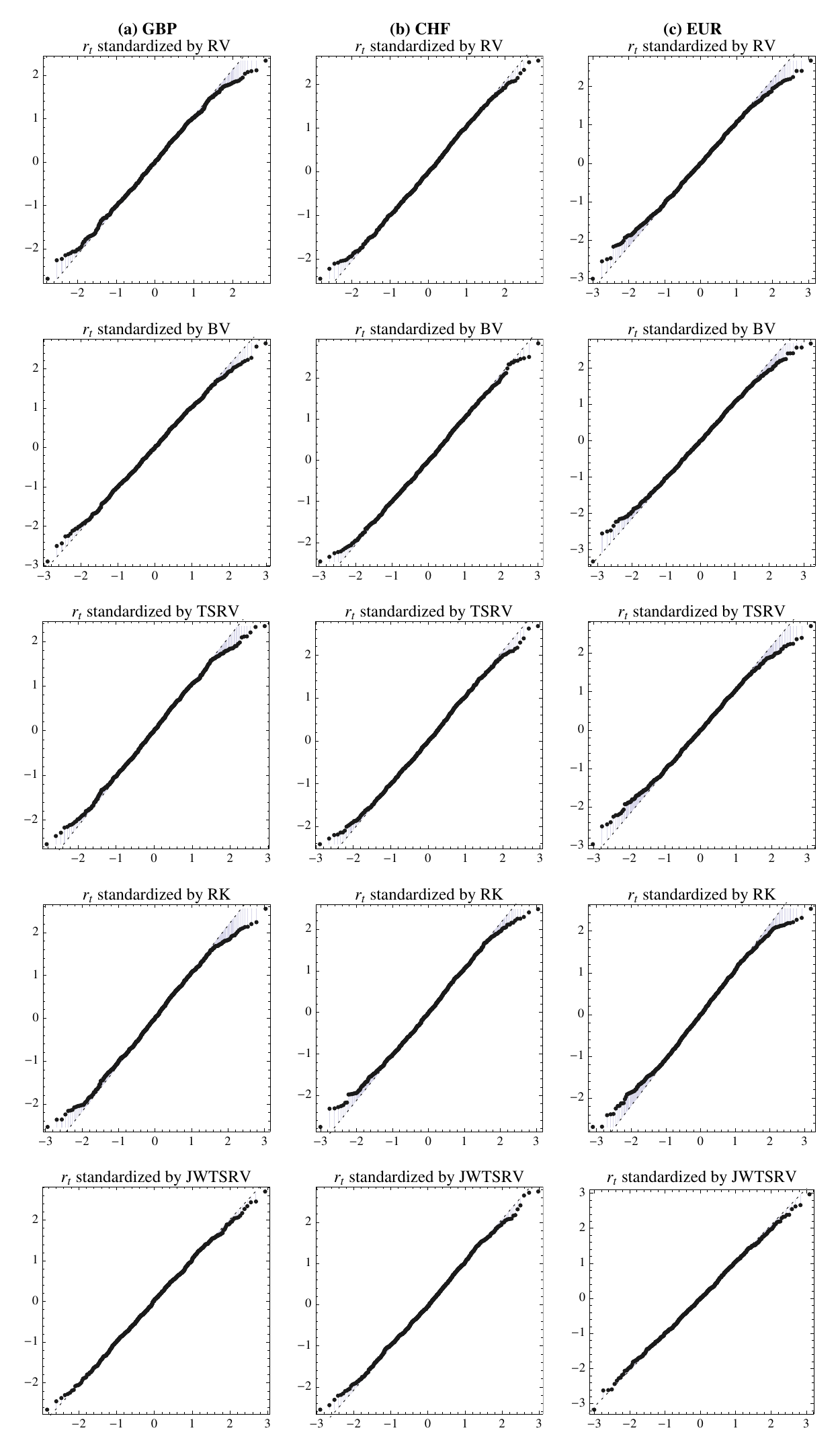} 
   \caption{QQ plots of normalized daily log-returns $r_t$ by RV, BV, TSRV, RK and JWTSRV estimators. (a) GBP futures, (b) CHF futures and (c) EUR futures}
   \label{qqplots}
\end{figure}

Table \ref{summarystats2} summarizes the unconditional distribution of the daily log-returns standardized by the integrated volatility, $r_t/IV_t^{1/2}$, and confirms this result. However, quite significant differences can be found among the estimators. While the high kurtosis (above 4) for the raw returns is reduced to the range of 2.51--2.81 for the log-returns standardized using the integrated volatility estimator, there is a notable difference between the estimators. The RV is expected to perform the worst, as it should be biased by microstructure noise and jumps, which is confirmed. The TSRV as well as the RK are not biased by noise, but it still contains a jump component of integrated variance. The BV should consistently estimate the jump components; the statistical distribution of $r_t/IV_t^{1/2}$, where $IV_t$ is estimated by the BV, should be closer to Gaussian. Finally, we expect JWTSRV estimator to perform the best, as it proved to be robust to noise and jumps in the Monte Carlo simulations. We also borrow the QQ plots plotted in Figure \ref{qqplots} for help. Similarly as \cite{flemingpaye2011} and \cite{ABH2011}, we ask whether the jumps account for the non-normality of the unconditional log-returns standardized by the integrated volatility estimators found in the literature. We add the TSRV, RK and JWTSRV estimators for comparison. Figure \ref{qqplots} shows that returns standardized by integrated volatility using the JWTSRV provide the best approximation of the standard normal distribution. This result is in line with what we expected, as the JWTSRV proved to be robust to noise and jumps in our large Monte Carlo study. The result from the BV leaves us puzzled. While it is expected to be robust to jumps, it should be able to perform better. The returns standardized by the BV have higher kurtosis than those standardized by the RV, TSRV or RK, thus the BV outperforms these estimators to some extent. However, the JWTSRV confirms the theory presented in the previous sections.

Moving from the distributional properties of the standardized daily log-returns, Table \ref{summarystats2} also shows the distributional properties of the $IV_t^{1/2}$ estimators. Again, the JWTSRV provides lower estimates of $IV_t^{1/2}$ and is also less volatile than the RV. This finding is consistent with the fact that the RV can be affected by microstructure noise, and, as demonstrated in the Monte Carlo simulations, the JWTSRV is able to estimate the true integrated variance with the lowest bias in the presence of noise and jumps in the data. It is surprising, though, that the average estimate of $IV_t^{1/2}$ using the JWTSRV is 6.34\% lower than the average estimate from the RV (computed as arithmetic averages on the estimators on GBP futures, CHF futures and EUR futures) with kurtosis 12.32\% lower than the RV. The average  estimate of $IV_t^{1/2}$ using the JWTSRV is 3.76\% lower than the average estimate using GBP, with kurtosis 6.34\% lower. The average  estimate of $IV_t^{1/2}$ using the JWTSRV is 4.52\% lower than the average estimate using the TSRV, with kurtosis 4.39\% lower. Finally, the average estimate of $IV_t^{1/2}$ using the JWTSRV is the same as the average estimate using the RK with kurtosis 25.39\% lower. It is thus interesting that while the TSRV accounts for noise but not jumps and the BV accounts for jumps but is not able to deal with noise, they have same deviations from the JWTSRV, which seems to estimate the integrated volatility without jumps and noise. Most interesting is that the average estimate of the RK is exactly the same as the average estimate of the JWTSRV. However, the RK estimates has much higher kurtosis. This result shows that the RK is powerful estimator of the realized variance. Finally let us note that these differences are economically significant, as they result in different asset pricing.

\subsection{$IV_t$ decomposition using wavelets \label{decompositionofIV}}
From the numerical analysis, we could see that the JWTSRV provides a feasible estimator of integrated variance. Another advantage is that by using our estimator, we are able to decompose the integrated  variance into several investment horizons, or components. In our analysis, we limit ourselves\footnote{It should be noted that any investment horizons of interest may be chosen arbitrarily.} to decomposition into four scales corresponding to investment horizons of 10 minutes, 20 minutes, 40 minutes and 80 minutes, and the rest up to 1 day. As shown in the theoretical part of this work, we can comfortably decompose the integrated variance into these components, as their sum will always give the integrated variance estimator. 

%
%

\begin{figure}[t]
   \centering
   \includegraphics[width=\textwidth]{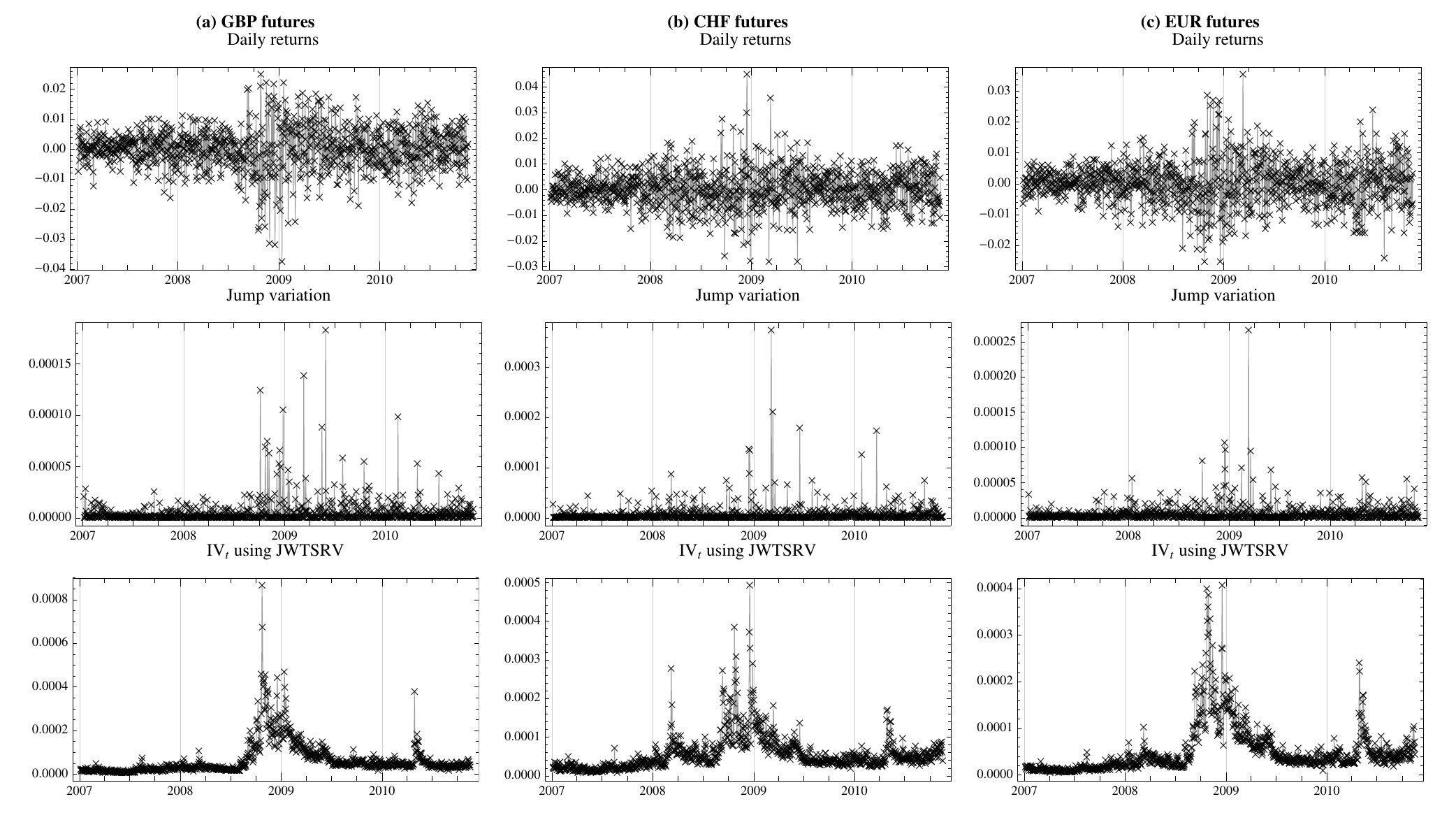} 
   \caption{Daily returns, estimated jump variation and $IV_t$ estimated by JWTSRV for (a) GBP, (b) CHF and (c) EUR futures.}
   \label{plotsBP}
\end{figure}

More precisely, the components of the $\widehat{RV}_{j,t,h}^{(JWTSRV)}$ from Eq. (\ref{jwtsrv}) correspond to various investment horizons. Thus, we will refer to these as $JWTSRV_j$, where $j=1,\dots,5$ are components corresponding to 10 minutes, 20 minutes, 40 minutes, 80 minutes and the rest up to 1 day investment horizon. 

\begin{figure}
   \centering
   \includegraphics[width=6.3in]{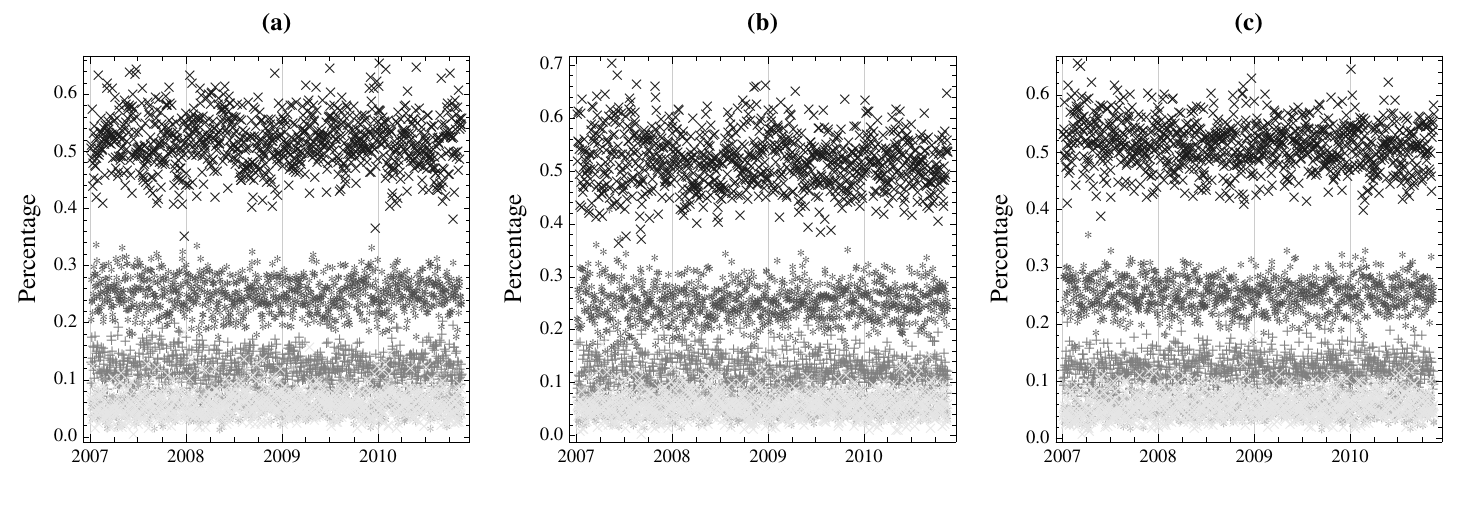} 
  \caption{ Contributions of components of integrated volatility $IV_t$ corresponding to investment horizons of 10 minutes (``$\times$" in black), 20 minutes (``$*$" in black), 40 minutes (``$+$" in grey), 80 minutes (``$\times$" light in grey) and up to 1 day (``$*$" in light grey). (a) GBP futures, (b) CHF futures and (c) EUR futures.}
  \label{plotebergies1}
\end{figure}

\begin{figure}
   \centering
   \includegraphics[width=6.3in]{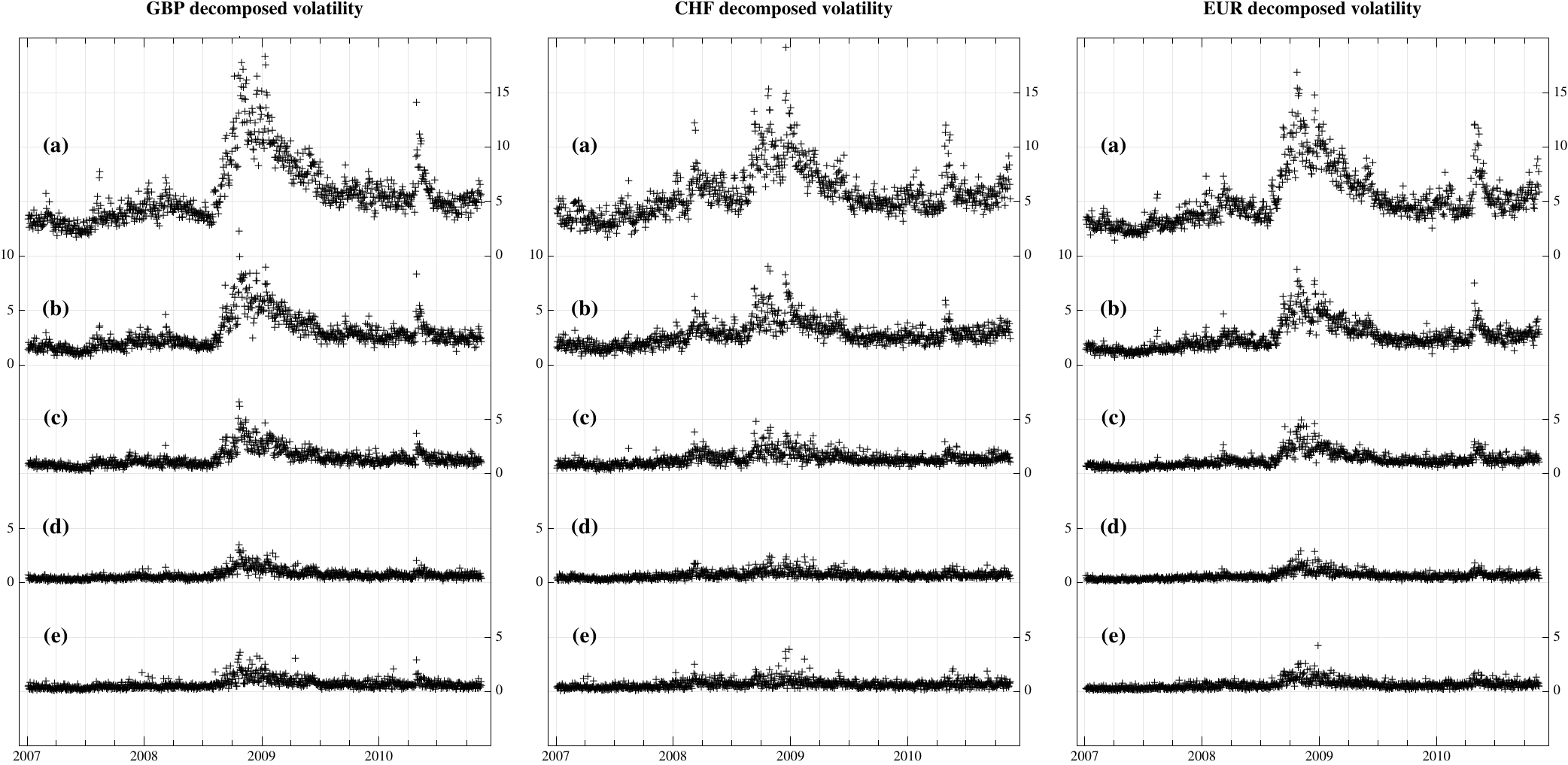} 
   \caption{Decomposed annualized volatility (by 252 days). (a) total $IV_{t}^{1/2}$ estimate on GBP, CHF and EUR futures using JWTSRV, (b) volatility on investment horizon of 10 minutes, (c) volatility on investment horizon of 20 minutes, (d) volatility on investment horizon of 40 minutes, (e) volatility on investment horizon of 80 minutes, (f) volatility on investment horizon of 1 day. Note that sum of components (b), (c), (d), (e) and (f) give total volatility plotted in (a).} 
   \label{volatilityplot}
\end{figure}

The decomposition of volatility into the so-called continuous and jump part is depicted by Figure \ref{plotsBP}, which provide the returns, estimated jumps and finally integrated variances using JWTSRV estimator for all three futures pairs. Figure \ref{volatilityplot} shows the further decomposition into several investment horizons. For better illustration, we annualize the square root of the integrated variance in order to get the annualized volatility and we compute the components of the volatility on several investment horizons. Figure \ref{volatilityplot} (a) to (e) show  the investment horizons of 10 minutes, 20 minutes, 40 minutes, 80 minutes and up to 1 day, respectively. It is very interesting that most of the volatility (around 50\%) comes from the fast, 10-minute investment horizon which is a new insight. In fact, it is a logical finding, as it shows that volatility is created on fast scales of up to 10 minutes rather than on slower scales. The longer the horizon, the lower the contribution of the variance to the total variation. We compute the weighted contributions of various investment horizon volatilities to the total to see its dynamics in time. More precisely, we compute the contributions of each scale to total variation  as:
\begin{equation}
\label{JWTSRVratio}
\widehat{RV}_{j,t}^{(JWTSRV)} / \widehat{RV}_{t}^{(JWTSRV)},
\end{equation}
for each $j=1,\dots,5$. The results are shown in Figure \ref{plotebergies1} for all investment horizons. Ratio in Eq. (\ref{JWTSRVratio}) is intuitive. If it equals zero, the investment horizon $j$ has zero contribution to the overall variance. If it equals one, the corresponding investment horizon $j$ explains all of the total variance. From Figure \ref{plotebergies1} we can see that the ratios are the same through all the currencies tested. They change quite considerably over the sample period. While the contribution of the first investment horizon, $j=1$, corresponding to 10 minutes investment horizon, to the total $IV_t$ is around 51.5\%, it is also the one with the largest dispersion. Over time, it changes from 40\% to 60\%. The second investment horizon (20 minutes), corresponding to $j=2$, accounts for approximately 25\% of the variance, followed by the third and fourth horizons (40 and 80 minutes, corresponding to $j=3$ and $j=4$), which account for only 12\% and 6\% approximately. The remaining 5\%--6\% are in the last $j=5$.

\section{Conclusion}
In this paper, we present the wavelet-based methodology for estimation of realized variance allowing its measurement in the time-frequency domain. To support our estimator, a numerical study of the finite sample performance of the estimator is carried out. In this study, we compare our estimator to several of the most popular estimators, namely, realized variance, bipower variation, two-scale realized volatility and realized kernels. The wavelet-based estimator proves to have lowest bias of all the estimators in the jump-diffusion model with stochastic volatility as well as the fractional stochastic volatility model simulated with different levels of noise and numbers of jumps. While all the other estimators suffer from substantial bias caused either by jumps or by noise, our theory proves to hold its properties under both noise and jumps.  As predictability of volatility is of interest to researchers as well as practitioners, a numerical study of the behavior of the forecasts is also carried out. Again, our theory proves to be the most powerful in forecasting volatility under the different simulation settings.

In addition, we use the estimator to decompose the empirical volatility and study its behavior at several different investment horizons. By studying the statistical properties of unconditional daily log-return distributions standardized by volatility estimated using the different estimators we find that standardization by our wavelet-based estimator brings the returns close to the Gaussian normal distribution. All the other estimators are affected by the presence of jumps in the data. The differences are large, as we find that the average volatility estimated using our wavelet-based estimator is 6.34\% lower than the volatility estimated with the standard estimators. 

Concluding the empirical findings, we show that our wavelet-based estimator brings a significant improvement to volatility estimation while it offers a time-frequency way of realized volatility measurement which helps us to better understand the dynamics of stock market behavior. Specifically, our theory uncovers that most of the volatility is created on higher frequencies.

\appendix


\section{Wavelets introduction}
\label{Appwaveintro}

This appendix briefly introduces the continuous wavelet transform (CWT) and maximal overlap discrete wavelet transformation (MODWT) needed for understanding of the proposed estimators in this paper.

Firstly, we outline the theoretical framework of continuous wavelet transform (CWT). 
\begin{def1} Continuous wavelet transform \citep{Daubechies1988}\\  
If $\psi\in\EL$ satisfies the admissibility condition
\begin{equation}
C_{\psi}:= \int_{\R} \left\vert \hat{\psi}(s)\right\vert^2\frac{1}{\vert s\vert}\, \mathrm{d}s\,<\,+\infty,
\end{equation}
where  $\hat{}$ denotes the Fourier transform, then $\psi$  is called a basic wavelet. 
Relative to every basic wavelet $\psi$, the continuous (integral) transform on $\EL$ is defined by
\begin{equation}
(W_\psi f)(j,k)=\langle\psi_{j,k},f\rangle=\mid j \mid ^{-1/2}\int_{\mathbb{R}}\overline{\psi\left(\frac{s-k}{j}\right)} f(s) \mathrm{d}s \hspace{5mm} f\in L^2(\mathbb{R}),
\end{equation}
where $\langle. ,.\rangle$ defines the $L^2$-inner product and $j,k\,\in\R$ with $j\neq 0$.
\end{def1}

Next we introduce the Calder\'on reconstruction formula \citep{chui1992}.

\begin{prop} Calder\'on reconstruction formula \\
\label{Calderon} 
Let $\psi\in\EL$ be a basic  wavelet which defines a continuous wavelet transform $\left(W_{\psi}f\right) (j,k)$. Then for any $f\in \EL$ and $s\in \R$ at which $f$ is continuous,
\begin{equation}
f(s)=\frac{1}{C_\psi}  \int_{\R}\int_{\R} (W_\psi f)(j,k) \psi_{j,k}(s)  \frac{1}{j^2}\mathrm{d}k \mathrm{d}j.
\end{equation}\\
Furthermore, let   $\psi$  satisfy the extra conditions
\begin{equation}
 \int_{0}^{+\infty} \left\vert \hat{\psi}(s)\right\vert^2\frac{1}{s}\, \mathrm{d}s=\int_{0}^{+\infty} \left\vert \hat{\psi}(-s)\right\vert^2\frac{1}{s}\, \mathrm{d}s=\frac{1}{2} C_{\psi}.
\end{equation}
Then
\begin{equation}
f(s)=\frac{2}{C_{\psi}}\int_0^{+\infty}\left[\int_{\R}\left(W_{\psi}f\right)(j,k)\psi_{j,k}(s)\,\mathrm  {d}k\right]\frac{1}{j^2}\, \mathrm{d} j
\end{equation}
for any $f\in \EL$ and $s\in \R$ at which $f$ is continuous.
\end{prop}
For the proof, see \cite{chui1992}.\\

Here we introduce the admissibility condition ensures that the Fourier transform of the wavelet $\hat{\psi}(s)$ has sufficient decay as $s\rightarrow 0$ \citep{Daubechies1988}. The finiteness of $C_\psi$ is guaranteed if $\hat{\psi}(0)=0$, which is equivalent to zero mean of the wavelet $\psi(.)$ \citep{Mallat98},
\begin{equation}
\label{c1}
\hat{\psi}(0)=\int_{-\infty}^\infty\psi(s)ds=0.
\end{equation}
Further, we impose the unit energy condition on the wavelet $\psi(.)$
\begin{equation}
\label{c2}
\int_{-\infty}^\infty\vert\psi(s)\vert^2 ds=1
\end{equation}

Conditions \ref{c1} and \ref{c2} ensure that the wavelet has some non-zero terms, but all excursions away from zero must cancel out. Detailed discussion about the wavelets and wavelet transform can be found in \cite{Daubechies1988}, \cite{Daubechies1992} and \cite{Gencay2002}.\\

\subsection{The maximal overlap discrete wavelet transform}
\label{modwt}

The maximal overlap discrete wavelet transformation (MODWT) is a special type discrete wavelet transform that is not sensitive to the choice of starting point of the examined time series, i.e. it is translation-invariant transform. Unlike the discrete wavelet transform \citep{Mallat98} (DWT), the MODWT does not use a downsampling procedure, therefore, the wavelet and scaling coefficient vectors at all scales have equal length \citep{PercivalWalden2000, Gencay2002}. As a consequence, the sample size of the examined process is not restricted to the powers of two, which makes the transform very useful for empirical data analysis. 

The MODWT wavelet and scaling coefficients can be conveniently used for an analysis of variance of stochastic processes in the time-frequency domain. \cite{Percival1995} clearly shows the advantages of the MODWT estimator of variance over the DWT estimator. Moreover, \cite{Serroukh2000} analyze the statistical properties of the MODWT variance estimator for non-stationary and non-Gaussian processes. For more details about the MODWT see \cite{Mallat98}, \cite{PercivalWalden2000} and \cite{Gencay2002}.

\subsubsection{Definition of MODWT filters}

We begin our description of the MODWT with introduction of wavelet filters. In a wavelet transform we use two types of filters; the scaling filter which is a low-pass filter and the wavelet filter that is a high-pass filter. The MODWT scaling and wavelet filters denoted as $g_l$ and $h_l$, have length $L$, $l=0,1,\ldots,L-1$. In our analysis we use the Daubechies D(4) wavelet filter with the filter length $L=4$ \citep{Daubechies1992}. There are three basic properties that both the MODWT filters must fulfill. Let us show these properties for the MODWT wavelet filter:

\begin{equation}
\sum_{l=0}^{L-1}h_{l}=0,\hspace{2mm}  \sum_{l=0}^{L-1}h_{l}^2=1/2, \hspace{2mm}  \sum_{l=-\infty}^{\infty} h_{l} h_{l+2N}=0, \hspace{2mm} N\in\ZN,
\end{equation}
then specifically for the MODWT scaling filter:
\begin{equation}
\sum_{l=0}^{L-1}g_{l}=1,\hspace{2mm}  \sum_{l=0}^{L-1}g_{l}^2=1/2, \hspace{2mm}  \sum_{l=-\infty}^{\infty} g_{l} g_{l+2N}=0, \hspace{2mm} N\in\ZN.
\end{equation}

\subsubsection{Pyramid algorithm}

With appropriate wavelet and filters we can proceed to compute the wavelet and scaling coefficients.  We obtain the MODWT wavelet and scaling coefficients using the pyramid algorithm \citep{Mallat98, PercivalWalden2000}. The first scale wavelet coefficients ($j=1$) are computed via filtering the process $x_t$ for $t=1,\ldots,N$ with the MODWT wavelet and scaling filters: 
\begin{equation}
W_{1,k}\equiv\sum_{l=0}^{L-1}h_{l}x_{k-l mod N}, \hspace{5mm} V_{1,k}\equiv\sum_{l=0}^{L-1}g_{l}x_{k-l mod N}.
\end{equation}

In the second stage of the pyramid algorithm, we replace $x_t$ with the scaling coefficients $V_{1,k}$ and after the filtering we obtain wavelet coefficients at the second scale $j=2$ as:
\begin{equation}
W_{2,k}\equiv\sum_{l=0}^{L-1}h_{l}V_{1,k-l mod N}, \hspace{5mm} V_{2,k}\equiv\sum_{l=0}^{L-1}g_{l}V_{1,k-l mod N}.
\end{equation}
We can proceed similar way to get The $j$-th level MODWT coefficients are in the form: 
\begin{equation}
W_{j,k}\equiv\sum_{l=0}^{L-1}h_{l}V_{j-1,k-l mod N}, \hspace{5mm} V_{j,k}\equiv\sum_{l=0}^{L-1}g_{l}V_{j-1,k-l mod N}, \hspace{3mm} j=1,2,\ldots, J^m.
\end{equation}
where $J^m\le log_2(N)$ is the maximum level of decomposition. Vector of MODWT coefficients wavelet $\mathbf{W}_1$ represents the frequency band $f\in[1/4,1/2]$, $\mathbf{W}_2$: $f\in[1/8,1/4]$ and $\mathbf{V}_2$: $f\in[0,1/8]$. The $j$-th level wavelet coefficients in the vector $\mathbf{W}_j$ represents frequency bands $f\in[1/2^{j+1},1/2^{j}]$ whereas the $j$-th level scaling coefficients in the vector $\mathbf{V}_j$ represents $f\in[0,1/2^{j+1}]$. For our estimator we use the MODWT wavelet coefficients that are unaffected by the boundary conditions. 

Finally, we define a vector $\mathbf{\mathcal{W}}$ consisting of $J^m+1$ subvectors of dimensions $N$, where the first $J^m$ subvectors are the MODWT wavelet coefficients at levels $j=1,...,N$ and the last subvector is the MODWT scaling coefficients at a level $J^m$:
\begin{equation}
\label{defW}
\mathcal{W}=\left[ \mathbf{W}_1,\mathbf{W}_2, \ldots, \mathbf{W}_{J^m}, \mathbf{V}_{J^m}\right]^T.
\end{equation}

\bibliography{thesis}

\begin{thebibliography}{}

\bibitem[\protect\citeauthoryear{A\"{i}t-Sahalia and Jacod}{A\"{i}t-Sahalia and
  Jacod}{2009}]{sahaliajacod2009}
A\"{i}t-Sahalia, Y. and J.~Jacod (2009).
\newblock Testing for jumps in a discretely observed process.
\newblock {\em The Annals of Statistics\/}~{\em 37\/}(1), 184--222.

\bibitem[\protect\citeauthoryear{A\"{i}t-Sahalia and Mancini}{A\"{i}t-Sahalia
  and Mancini}{2008}]{Sahalia2008}
A\"{i}t-Sahalia, Y. and L.~Mancini (2008).
\newblock Out of sample forecasts of quadratic variation.
\newblock {\em Journal of Econometrics\/}~(147), 17--33.

\bibitem[\protect\citeauthoryear{Andersen and Bollerslev}{Andersen and
  Bollerslev}{1998}]{ab98}
Andersen, T. and T.~Bollerslev (1998).
\newblock Answering the skeptics: Yes, standard volatility models do provide
  accurate forecasts.
\newblock {\em International Economic Review\/}~(39).

\bibitem[\protect\citeauthoryear{Andersen, Bollerslev, Diebold, and
  Ebens}{Andersen et~al.}{2001}]{abde2001}
Andersen, T., T.~Bollerslev, F.~Diebold, and H.~Ebens (2001).
\newblock The distribution of realized stock return volatility.
\newblock {\em Journal of Financial Economics\/}~{\em 61}, 43--76.

\bibitem[\protect\citeauthoryear{Andersen, Bollerslev, Diebold, and
  Labys}{Andersen et~al.}{2001}]{abdl2001}
Andersen, T., T.~Bollerslev, F.~Diebold, and P.~Labys (2001).
\newblock The distribution of realized exchange rate volatility.
\newblock {\em Journal of the American Statistical Association\/}~(96), 42--55.

\bibitem[\protect\citeauthoryear{Andersen, Bollerslev, Diebold, and
  Labys}{Andersen et~al.}{2003}]{abdl2003}
Andersen, T., T.~Bollerslev, F.~Diebold, and P.~Labys (2003).
\newblock Modeling and forecasting realized volatility.
\newblock {\em Econometrica\/}~(71), 579--625.

\bibitem[\protect\citeauthoryear{Andersen, Bollerslev, and Huang}{Andersen
  et~al.}{2011}]{ABH2011}
Andersen, T., T.~Bollerslev, and X.~Huang (2011).
\newblock A reduced form framework for modeling volatility of speculative
  prices based on realized variation measures.
\newblock {\em Journal of Econometrics\/}~{\em 160\/}(1), 176--189.

\bibitem[\protect\citeauthoryear{Andersen, Bollerslev, and Diebold}{Andersen
  et~al.}{2007}]{abd2007}
Andersen, T.~G., T.~Bollerslev, and F.~X. Diebold (2007).
\newblock Roughing it up: Including jump components in the measurement,
  modeling, and forecasting of return volatility.
\newblock {\em Review of Economics and Statistics\/}~(4), 701--720.

\bibitem[\protect\citeauthoryear{Andersen, Dobrev, and Schaumburg}{Andersen
  et~al.}{2009}]{ads2009}
Andersen, T.~G., D.~Dobrev, and E.~Schaumburg (2009).
\newblock Jump-robust volatility estimation using nearest neighbor truncation.
\newblock {\em Working paper, National Bureau of Economic Research\/}.

\bibitem[\protect\citeauthoryear{Antoniou and Gustafson}{Antoniou and
  Gustafson}{1999}]{antoniou1999}
Antoniou, I. and K.~Gustafson (1999).
\newblock Wavelets and stochastic processes.
\newblock {\em Mathematics and Computers in Simulation\/}~{\em 49}, 81--104.

\bibitem[\protect\citeauthoryear{Bandi and Russell}{Bandi and
  Russell}{2006}]{bandirussel2006a}
Bandi, F. and J.~Russell (2006).
\newblock Separating microstructure noise from volatility.
\newblock {\em Journal of Financial Economics\/}~(79), 655--692.

\bibitem[\protect\citeauthoryear{Barndorff-Nielsen, Hansen, Lunde, and
  Shephard}{Barndorff-Nielsen et~al.}{2008}]{barndorff2008}
Barndorff-Nielsen, O., P.~Hansen, A.~Lunde, and N.~Shephard (2008).
\newblock Designing realized kernels to measure the ex-post variation of equity
  prices in the presence of noise.
\newblock {\em Econometrica\/}~{\em 76\/}(6), 1481--1536.

\bibitem[\protect\citeauthoryear{Barndorff-Nielsen and
  Shephard}{Barndorff-Nielsen and Shephard}{2001}]{barndorff2001}
Barndorff-Nielsen, O. and N.~Shephard (2001).
\newblock Non-gaussian ornstein-uhlenbeck-based models and some of their uses
  in financial economics.
\newblock {\em Journal of the Royal Statistical Society, Series B\/}~(63),
  167--241.

\bibitem[\protect\citeauthoryear{Barndorff-Nielsen and
  Shephard}{Barndorff-Nielsen and Shephard}{2002a}]{barndorff2002a}
Barndorff-Nielsen, O. and N.~Shephard (2002a).
\newblock Econometric analysis of realised volatility and its use in estimating
  stochastic volatility models.
\newblock {\em Journal of the Royal Statistical Society, Series B\/}~(64),
  253--280.

\bibitem[\protect\citeauthoryear{Barndorff-Nielsen and
  Shephard}{Barndorff-Nielsen and Shephard}{2002b}]{barndorff2002}
Barndorff-Nielsen, O. and N.~Shephard (2002b).
\newblock Estimating quadratic variation using realized variance.
\newblock {\em Journal of Applied Econometrics\/}~(17), 457--477.

\bibitem[\protect\citeauthoryear{Barndorff-Nielsen and
  Shephard}{Barndorff-Nielsen and Shephard}{2004}]{barndorff2004}
Barndorff-Nielsen, O. and N.~Shephard (2004).
\newblock Power and bipower variation with stochastic volatility and jumps.
\newblock {\em Journal of Financial Econometrics\/}~(2), 1--48.

\bibitem[\protect\citeauthoryear{Barndorff-Nielsen and
  Shephard}{Barndorff-Nielsen and Shephard}{2006}]{barndorff2006}
Barndorff-Nielsen, O. and N.~Shephard (2006).
\newblock Econometrics of testing for jumps in financial economics using
  bipower variation.
\newblock {\em Journal of Financial Econometrics\/}~(4), 1--30.

\bibitem[\protect\citeauthoryear{Capobianco}{Capobianco}{2004}]{capobianco2004}
Capobianco, E. (2004).
\newblock Multiscale stochastic dynamics in finance.
\newblock {\em Physica A\/}~(344), 122--127.

\bibitem[\protect\citeauthoryear{Chui}{Chui}{1992}]{chui1992}
Chui, C. (1992).
\newblock {\em An Inroduction to Wavelets}.
\newblock Academic Press, New York.

\bibitem[\protect\citeauthoryear{Comte and Renault}{Comte and
  Renault}{1999}]{comte98}
Comte, F. and E.~Renault (1999).
\newblock Long memory in continuous-time stochastic volatility models.
\newblock {\em Mathematical Finance\/}~(8), 291--323.

\bibitem[\protect\citeauthoryear{Daubechies}{Daubechies}{1988}]{Daubechies1988}
Daubechies, I. (1988).
\newblock Orthonormal bases of compactly supported wavelets.
\newblock {\em Communications on Pure and Applied Mathematics\/}~{\em 41},
  909--996.

\bibitem[\protect\citeauthoryear{Daubechies}{Daubechies}{1992}]{Daubechies1992}
Daubechies, I. (1992).
\newblock {\em Ten lectures on wavelets}.
\newblock SIAM.

\bibitem[\protect\citeauthoryear{Donoho and Johnstone}{Donoho and
  Johnstone}{1994}]{donoho}
Donoho, D.~L. and I.~M. Johnstone (1994).
\newblock Ideal spatial adaptation by wavelet shrinkage.
\newblock {\em Biometrica\/}~(81), 425--455.

\bibitem[\protect\citeauthoryear{Fama}{Fama}{1965}]{fama}
Fama, E. (1965).
\newblock The behavior of stock market prices.
\newblock {\em Journal of Business\/}~{\em 38}, 34--105.

\bibitem[\protect\citeauthoryear{Fan and Wang}{Fan and
  Wang}{2007}]{fanwang2008}
Fan, J. and Y.~Wang (2007).
\newblock Multi-scale jump and volatility analysis for high-frequency financial
  data.
\newblock {\em Journal of the American Statistical Association\/}~(102),
  1349--1362.

\bibitem[\protect\citeauthoryear{Fleming and Paye}{Fleming and
  Paye}{2011}]{flemingpaye2011}
Fleming, J. and B.~Paye (2011).
\newblock High-frequency returns, jumps and the mixture of normals hypothesis.
\newblock {\em Journal of Econometrics\/}~{\em 160\/}(1), 119--128.

\bibitem[\protect\citeauthoryear{Gen{\c c}ay, Sel{\c c}uk, and Whitcher}{Gen{\c
  c}ay et~al.}{2002}]{Gencay2002}
Gen{\c c}ay, R., F.~Sel{\c c}uk, and B.~Whitcher (2002).
\newblock {\em An Introduction to Wavelets and Other Filtering Methods in
  Finance and Economics.}
\newblock Academic Press.

\bibitem[\protect\citeauthoryear{Gen\c{c}ay, Gradojevic, Sel\c{c}uk, and
  Whitcher}{Gen\c{c}ay et~al.}{2010}]{gencay2010}
Gen\c{c}ay, R., N.~Gradojevic, F.~Sel\c{c}uk, and B.~Whitcher (2010).
\newblock Asymmetry of information flow between volatilities across time
  scales.
\newblock {\em Quantitative Finance\/}, 1--21.

\bibitem[\protect\citeauthoryear{Hansen and Lunde}{Hansen and
  Lunde}{2006}]{hansenlunde2006}
Hansen, P. and A.~Lunde (2006).
\newblock Realized variance and market microstructure noise.
\newblock {\em Journal of Business and Economic Statistics\/}~{\em 24\/}(2),
  127--161.

\bibitem[\protect\citeauthoryear{H\o~g and Lunde}{H\o~g and
  Lunde}{2003}]{hoglunde2003}
H\o~g, E. and A.~Lunde (2003).
\newblock Wavelet estimation of integrated volatility.
\newblock {\em Working Paper. Aarhus School of Business.\/}.

\bibitem[\protect\citeauthoryear{Huang and Tauchen}{Huang and
  Tauchen}{2005}]{huang2005}
Huang, X. and G.~Tauchen (2005).
\newblock The relative contribution of jumps to total price variance.
\newblock {\em Journal of Financial Econometrics\/}~{\em 3}, 456--499.

\bibitem[\protect\citeauthoryear{Mallat}{Mallat}{1998}]{Mallat98}
Mallat, S. (1998).
\newblock {\em A wavelet tour of signal processing}.
\newblock Academic Press.

\bibitem[\protect\citeauthoryear{Mancini}{Mancini}{2009}]{Mancini2009}
Mancini, C. (2009).
\newblock Non-parametric threshold estimation for models with stochastic
  diffusion coefficient and jumps.
\newblock {\em Scandinavian Journal of Statistics\/}~{\em 36}, 270--296.

\bibitem[\protect\citeauthoryear{Mancino and Sanfelici}{Mancino and
  Sanfelici}{2008}]{mancino2008}
Mancino, M. and S.~Sanfelici (2008).
\newblock Robustness of fourier estimator of integrated volatility in the
  presence of microstructure noise.
\newblock {\em Computational Statistics \& data analysis\/}~{\em 52\/}(6),
  2966--2989.

\bibitem[\protect\citeauthoryear{Mandelbrot}{Mandelbrot}{1963}]{mandelbrot}
Mandelbrot, B. (1963).
\newblock The variation of certain speculative prices.
\newblock {\em Journal of Business\/}~{\em 36}, 394--419.

\bibitem[\protect\citeauthoryear{Marinucci and Robinson}{Marinucci and
  Robinson}{1999}]{marinucci99}
Marinucci, D. and P.~Robinson (1999).
\newblock Alternative forms of fractional brownian motion.
\newblock {\em Journal of Statistical Planning and Inference\/}~(80), 111--122.

\bibitem[\protect\citeauthoryear{Mincer and Zarnowitz}{Mincer and
  Zarnowitz}{1969}]{mincerzarnowitz}
Mincer, J. and V.~Zarnowitz (1969).
\newblock {\em The evaluation of economic forecasts}.
\newblock New York: National Bureau of Economic Research.

\bibitem[\protect\citeauthoryear{Nielsen and Frederiksen}{Nielsen and
  Frederiksen}{2008}]{nielsenfrederiksen2008}
Nielsen, M. and P.~Frederiksen (2008).
\newblock Finite sample accuracy and choice of sampling frequency in integrated
  volatility estimation.
\newblock {\em Journal of Empirical Finance\/}~(15), 265--286.

\bibitem[\protect\citeauthoryear{Olhede, Sykulski, and Pavliotis}{Olhede
  et~al.}{2009}]{olhede2009}
Olhede, S., A.~Sykulski, and G.~Pavliotis (2009).
\newblock Frequency domain estimation of integrated volatility for ito
  processes in the presence of market-microstructure noise.
\newblock {\em Multiscale Modeling \& Simulation\/}~{\em 8\/}(2), 393--427.

\bibitem[\protect\citeauthoryear{Percival}{Percival}{1995}]{Percival1995}
Percival, D.~B. (1995).
\newblock On estimation of the wavelet variance.
\newblock {\em Biometrika\/}~{\em 82}, 619--631.

\bibitem[\protect\citeauthoryear{Percival and Mofjeld}{Percival and
  Mofjeld}{1997}]{PercivalMofjeld1997}
Percival, D.~B. and H.~Mofjeld (1997).
\newblock Analysis of subtidal coastal sea level fluctuations using wavelets.
\newblock {\em Journal of the American Statistical Association\/}~{\em
  92\/}(439), 886--880.

\bibitem[\protect\citeauthoryear{Percival and Walden}{Percival and
  Walden}{2000}]{PercivalWalden2000}
Percival, D.~B. and A.~T. Walden (2000).
\newblock {\em Wavelet Methods for Time series Analysis}.
\newblock Cambridge University Press.

\bibitem[\protect\citeauthoryear{Protter}{Protter}{1992}]{protter}
Protter, P. (1992).
\newblock {\em Stochastic integration and differential equations: A new
  approach}.
\newblock New York: Springer-Verlag.

\bibitem[\protect\citeauthoryear{Serroukh, Walden, and Percival}{Serroukh
  et~al.}{2000}]{Serroukh2000}
Serroukh, A., A.~T. Walden, and D.~B. Percival (2000).
\newblock Statistical properties and uses of the wavelet variance estimator for
  the scale analysis of time series.
\newblock {\em Journal of the American Statistical Association\/}~{\em 95},
  184--196.

\bibitem[\protect\citeauthoryear{Subbotin}{Subbotin}{2008}]{subbotin2008}
Subbotin, A. (2008).
\newblock A multi-horizon scale for volatility.
\newblock Technical report, Documents de travail du Centre d'Economie de la
  Sorbonne, Universit{\'e} Panth{\'e}on-Sorbonne (Paris 1).

\bibitem[\protect\citeauthoryear{Wang}{Wang}{1995}]{wang95}
Wang, Y. (1995).
\newblock Jump and sharp cusp detection via wavelets.
\newblock {\em Biometrika\/}~(82), 385--397.

\bibitem[\protect\citeauthoryear{Zhang, Mykland, and A\"{i}t-Sahalia}{Zhang
  et~al.}{2005}]{zhang2005}
Zhang, L., P.~Mykland, and Y.~A\"{i}t-Sahalia (2005).
\newblock A tale of two time scales: Determining integrated volatility with
  noisy high frequency data.
\newblock {\em Journal of the American Statistical Association\/}~(100),
  1394--1411.

\end{thebibliography}
\bibliographystyle{chicago}

\end{document}